\documentclass[a4paper,12pt]{article}
\usepackage{jheppub,esint,shuffle,psfrag}
\usepackage[utf8]{inputenc}
\usepackage{slashed}

\usepackage{epsfig,amssymb,amsmath,psfrag,subfigure,rotate,color,wasysym,xcolor}
\usepackage[bbgreekl]{mathbbol}

\allowdisplaybreaks
\newcommand{\insertfig}[2]{\includegraphics[width=#1cm]{#2}}

\DeclareSymbolFontAlphabet{\mathbbm}{bbold}
\DeclareSymbolFontAlphabet{\mathbb}{AMSb}%

\def\XXint#1#2#3{{\setbox0=\hbox{$#1{#2#3}{\int}$ }
\vcenter{\hbox{$#2#3$ }}\kern-.6\wd0}}

\def \be  {\begin{equation}}
\def \ee  {\end{equation}}
\def \ba  {\begin{eqnarray}}
\def \ea  {\end{eqnarray}}
\def \baa {\begin{eqnarray*}}
\def \eaa {\end{eqnarray*}}
\def \lab #1 {\label{#1}}

\newcommand\re[1]{(\ref{#1})}
\def\d{\hbox{{d}\kern-.20em\hbox{l}}}
\def \qqquad {\qquad\quad}
\def \qqqquad {\qquad\qquad}
\def \matrix #1 {\left(\begin{array}{cc} #1 \end{array}\right)}

\def \tr {\mathop{\rm tr}\nolimits}

\def \e  {\mathop{\rm e}\nolimits}
\newcommand\lr[1]{{\left({#1}\right)}}

\newcommand \vev [1] {\langle{#1}\rangle}
\newcommand \VEV [1] {\left\langle{#1}\right\rangle}

\newcommand{\bit}[1]{\mbox{\boldmath$#1$}}
\def\1{\hbox{{1}\kern-.25em\hbox{l}}}

\newcommand{\ft}[2]{{\textstyle\frac{#1}{#2}}}

\makeatletter

\input pdf-trans
\newbox\qbox
\def\usecolor#1{\csname\string\color@#1\endcsname\space}
\newcommand\bordercolor[1]{\colsplit{1}{#1}}
\newcommand\fillcolor[1]{\colsplit{0}{#1}}
\newcommand\outline[1]{\leavevmode%
  \def\maltext{#1}%
  \setbox\qbox=\hbox{\maltext}%
  \boxgs{Q q 2 Tr \thickness\space w \fillcol\space \bordercol\space}{}%
  \copy\qbox%
}
\makeatother
\newcommand\colsplit[2]{\colorlet{tmpcolor}{#2}\edef\tmp{\usecolor{tmpcolor}}%
  \def\tmpB{}\expandafter\colsplithelp\tmp\relax%
  \ifnum0=#1\relax\edef\fillcol{\tmpB}\else\edef\bordercol{\tmpC}\fi}
\def\colsplithelp#1#2 #3\relax{%
  \edef\tmpB{\tmpB#1#2 }%
  \ifnum `#1>`9\relax\def\tmpC{#3}\else\colsplithelp#3\relax\fi
}
\bordercolor{black}
\fillcolor{white}
\def\thickness{.3}

\def\1{\mathbbm{1}}

\title{Circular Wilson loop in $\boldsymbol{\mathcal{N} = 2^\ast}$ super Yang-Mills theory at two loops and localization}
 
\author[a,b]{A.V.~Belitsky}
\author [b]{and G.P.~Korchemsky}
 \affiliation[a] {Department of Physics, Arizona State University, 
 Tempe, AZ 85287-1504, USA}

 \affiliation[b] {Institut de Physique Th\'eorique\footnote{Unit\'e Mixte de Recherche 3681 du CNRS}, Universit\'e Paris Saclay, CNRS, CEA, 91191 Gif-sur-Yvette} 
 
\preprint{  \parbox[t]{28mm}{IPhT--T20/011}}

 \abstract
{
We present a two-loop calculation of the supersymmetric circular Wilson loop in the $\mathcal{N} = 2^\ast$ super Yang-Mills theory on the four-sphere.
We develop an efficient framework for computing contributing Feynman graphs that relies on using the embedding coordinates combined with the 
Mellin-Barnes techniques for propagator-like integrals on the sphere. Our results exactly match predictions of supersymmetric localization providing a 
nontrivial consistency check for the latter in non-conformal settings.
 }

\begin{document}

\maketitle
\flushbottom
\setcounter{footnote} 0

\section{Introduction}

The supersymmetric localization offers a very powerful tool to study supersymmetric gauge theories both in perturbative and nonperturbative regimes, see, e.g., the review \cite{Pestun:2016zxk}.
It allows one to compute various observables in theories defined on the four-sphere in terms of interacting matrix models. One notable example of this is the $\mathcal{N} = 2^\ast$ 
theory, which arises as a massive deformation of the maximally supersymmetric Yang-Mills theory in four dimensions preserving $\mathcal{N} = 2$ supersymmetry.
The Lagrangian of the latter in conformally-flat spaces is uniquely fixed by its extended supersymmetry~\cite{Pestun:2007rz,Bobev:2013cja}. 
 
To date, the localization has been applied to calculations of partition functions, circular Wilson loops and certain correlators in various supersymmetric theories~\cite{Pestun:2016zxk}. 
The corresponding matrix model predictions have been successfully checked at weak coupling against explicit perturbative analyses in superconformal~\cite{Andree:2010na,Billo:2019fbi}  
and 
in massless non-conformal supersymmetric gauge theories
\cite{Baggio:2014sna,Baggio:2014ioa,Baggio:2015vxa,Gerchkovitz:2016gxx,Baggio:2016skg,Rodriguez-Gomez:2016ijh,Rodriguez-Gomez:2016cem,Billo:2017glv,Billo:2018oog,Billo:2019job}. 
In all of these circumstances, perturbation theory was developed in flat four-dimensional Euclidean space and observables were shown to be the same (at least to the order considered) as the 
ones on the four-sphere.

The use of flat-space analysis is invalidated, however, when the theory contains an additional scale, like a deformation mass-scale in the $\mathcal{N} = 2^\ast$ theory. In this case, 
conformal symmetry is explicitly broken and the theory possesses different properties at short and long distances. In particular, observables exhibit dependence on a dimensionless 
parameter determined by the product of the mass and the radius of the four-sphere. To check the localization formulas on the sphere, we have to therefore generalize  conventional Feynman 
integral technique to theories on curved backgrounds. This is what we are set to do in the present work.  Namely, we develop an appropriate formalism for evaluating Feynman diagrams 
on the sphere and apply it to computing the circular Wilson loop in the $\mathcal{N} = 2^\ast$ theory to two-loop order.

A convenient way to organize the calculation in the $\mathcal{N} = 2^\ast$ theory is to decompose all contributing graphs into those present in the $\mathcal N=4$ theory and remaining 
ones, as it was done in a related but different context in Refs.\ \cite{Andree:2010na,Billo:2017glv}. The contribution of the former diagrams is known explicitly whereas the latter define the 
difference between the circular Wilson loops in the two theories. To two-loop perturbative order, the difference only arises from graphs with gluon and scalar propagators dressed by 
one-loop corrections due to massive particles circulating in the virtual loop. This allows us to focus on propagator-like integrals containing massless external lines and 
massive loops.

Using the formalism of embedding coordinates~\cite{Adler:1972qq,Adler:1973ty,Drummond:1975yc,Drummond:1977uy}, it is straightforward to obtain a representation for two-loop corrections to 
the circular Wilson loop in terms of scalar integrals built from propagators of massless and massive particles. However, due to rather involved form of the latter, evaluating them by brute force 
becomes prohibitively complicated. We demonstrate in this paper, that all calculations can significantly be simplified by employing the Mellin-Barnes transformation for propagators involved. This 
yields a very compact representation for contributing graphs as triple Mellin-Barnes integrals. Expanding them at small and large values of the mass, we obtain expressions for the circular Wilson 
loop to two loops. We verify below that it agrees with the prediction from the localization and, at the same time, it differs from the analogous expression for the circular Wilson loop in flat space.

The rest of the paper is organized as follows. In Section \ref{sect:N=2}, we present the $\mathcal{N} = 2^\ast$ super Yang-Mills theory on the four-sphere. We then introduce the embedding 
coordinates, which prove to be of extreme value in our calculations, and provide expressions for free propagators of fields belonging to the vector and hypermultiplets. In Section 
\ref{SectionWilsonLoop}, we define the circular Wilson loop in $\mathcal{N} = 2^\ast$ theory. We compute it  to two loops in flat space and compare the result with the analogous expression 
on the sphere predicted by the supersymetric localization. In Section~\ref{SectionMB}, we use the embedding formalism to express relevant two-loop corrections in terms of scalar integrals 
on the sphere. The calculation of these integrals is discussed in Section~\ref{sect:comp}. We outline our results and check them against the localization prediction both numerically, for finite 
mass, and analytically, in the regime of small and large mass. In both cases we observe a perfect agreement. Concluding remarks are presented in Section~\ref{SectionConclusions}. Calculational 
details, which are too bulky for the main text, are summarized in six appendices. 

\section{$\boldsymbol {\mathcal{N} = 2^\ast}$ super Yang-Mills theory on the sphere}\label{sect:N=2}

In this section, we define the ${\mathcal{N} = 2^\ast}$ super Yang-Mills theory on the sphere. It arises as a mass deformation of the maximally supersymmetric $\mathcal N=4$ super Yang-Mills 
theory  in four-dimensional space. The deformation breaks the conformal symmetry of the latter but preserves the residual $\mathcal N=2$ Poincar\'e supersymmetry. The Lagrangian of this theory 
on the four-sphere has been constructed in Refs.~\cite{Pestun:2007rz,Bobev:2013cja} and we review it below.

\subsection{Mass deformation}

The maximally supersymmetric Yang-Mills theory can be obtained by compactifying the $\mathcal{N} = 1$ super Yang-Mills in ten space-time dimensions down to four spatial dimensions. 
It describes gauge fields $A_\mu$, six real scalars $X_I$ (with $I=1,\dots,6$) and four gauginos $\lambda^A$ and $\bar\lambda_A$ (with $A=1,\dots,4$) all belonging to the adjoint 
representation of the $SU(N)$ gauge group.~\footnote{The gauge fields and scalars arise as components of the gauge field $A_M = (A_\mu, X_I)$  in $D=10$ dimensional space-time with 
Minkowski signature. Rewriting the action of $\mathcal N=4$ theory in terms of these fields, one finds  that the last component of $A_M$ has a different sign in front of the kinetic term. This  
implies that the action is no longer real for conventionally defined involution of fields. This requires either compexification of the path integral or introduction of a different involution such as 
symplectic Majorana condition \cite{Belitsky:2000ii} or just giving up hermiticity altogether. }

The action of the $\mathcal N=4$ on the four-sphere $S^4$ looks as
\begin{align}\label{N4susy}
S_{\mathcal N=4} = \int d^4 x\, \sqrt{\mbox{\sl g}} \left({\mathcal L}_0 +{\mathcal L}_{\rm int}  \right)\,,
\end{align}
where $ {\mathcal L}_0$ contains minimally-coupled kinetic terms of the fields and ${\mathcal L}_{\rm int}$ describes Yukawa interaction and quartic self-interaction of scalars,
\begin{align}\notag\label{L}
 {\mathcal L}_0 &={\rm tr}\
\bigg[
\frac12 F^{\mu\nu} F_{\mu\nu}
- 2 i \lambda_{\alpha}^{A} \not\!\!\mathcal{D} ^{\alpha\beta'}
\bar\lambda_{\beta', A}
+ \frac12  \mathcal{D}^\mu \bar\phi_{AB} 
 \mathcal{D}_\mu \phi^{AB} + {1\over R^2} \bar\phi_{AB}\phi^{AB}\bigg]\,,
\\
{\mathcal L}_{\rm int} &={\rm tr}\
\bigg[  \sqrt{2} g\,
\bar\phi_{AB} \{\lambda^{\alpha, A}, \lambda_{\alpha}^{B}\}
+ \sqrt{2} g\,
\phi^{AB}
\{  \bar\lambda^{\alpha'}_{A}, \bar\lambda_{\alpha', B}\}
+ \frac{g^2}8 \left[ \phi^{AB}, \phi^{CD} \right] 
\left[ \bar\phi_{AB}, \bar\phi_{CD} \right] 
\bigg]\,.
\end{align}
Here the scalar fields are grouped into skew-symmetric matrices $\phi^{AB} =  \Sigma^{I, AB} X_{I}/\sqrt{2}$ and $\bar\phi_{AB} \equiv  \epsilon_{ABCD} \phi^{CD}/2$ (with $A,B=1,\dots,4$) 
with the expansion coefficients $\Sigma^{I\,, AB}$ being the chiral blocks of Dirac matrices in six dimensions. The gauginos $\lambda^A_\alpha$ and $\bar\lambda^{\alpha'}_A$ transform 
under the two subgroups of the rotation group in four dimensions $SO(4) = SU(2) \otimes SU'(2)$. The covariant derivative for gauginos looks as $\not\!\!\mathcal{D} ^{\alpha\beta'}\bar\lambda_{\beta', A}
= \slashed\nabla ^{\alpha\beta'}\bar\lambda_{\beta', A} - g [A^{\alpha\beta'}, \bar\lambda_{\beta', A}]$ where $\slashed\nabla ^{\alpha\beta'}$ is the spinor covariant derivative on the sphere 
(see Eq.~\re{nabla} below). Similar definition holds for derivatives acting on other fields, see, e.g., Eq.\ \re{LBoperator}. Finally, the last term in the expression for ${\mathcal L}_0$ describes the conformal 
coupling of the scalars to the scalar curvature of the four-sphere with radius $R$.

Upon the mass deformation, the massless $\mathcal N=4$ multiplet is decomposed into the $\mathcal N=2$ vector multiplet $(A,\lambda,\Phi )$ consisting of
\begin{align}
\label{VectorMultipletFields}
A_\mu \, , \qquad \lambda^1_\alpha \, , \qquad \lambda^2_\alpha \, , \qquad \Phi_1= X_3, \qquad \Phi_2= X_6
\, ,
\end{align}
and the massive hypermultiplet $(\psi,Z)$ consisting of
\begin{align}\label{HyperFields}
\psi^1_\alpha = \lambda^{3}_\alpha \, , \qquad \psi^2_\alpha = \lambda^{4}_\alpha \, , \qquad  Z_i= \frac{1}{\sqrt{2}} \left( X_i + i X_{i+3} \right) \, ,  \end{align}
where $i=1,2$.
It is straightforward to rewrite the action \re{N4susy}  in terms of these fields. To save space, we refrain from doing it. Also, the scalar field $\Phi_2$ does not appear in our calculation and we do not 
display it in all formulas that follow.

The mass deformation of the $\mathcal N=4$ theory is driven by the following term
\begin{align}\notag\label{mass}
{\mathcal L}_{\rm mass} &=  \tr \left[ 4 i g m   (\bar{Z}_1 Z_2-\bar{Z}_2 Z_1)  \Phi + 2 m^2 \bar{Z}_i Z_i\right] 
\\[2mm]
&  - i m \tr  \left( \psi^{i \alpha} \psi^i_\alpha + \bar\psi^{i\alpha'} \bar\psi_{i \alpha'} \right) + \frac{i m}{R}  \tr \left( Z_i^2 + \bar{Z}_i^2 \right)\,,
\end{align}
where $\bar{Z}_i$ and $\bar\psi^{i\alpha'}$ are conjugated fields to their unbarred counterparts and we displayed only the coupling to the real scalar $\Phi \equiv \Phi_1$.~\footnote{Strictly speaking, the 
fields that are complex conjugated in Lorentzian signature are  independent in Euclidean signature.} In flat space, i.e., for $R\to\infty$, it is generated by the superpotential \cite{Buchel:2000cn},
\begin{align}
W = \tr \left( 2 i g\, [Z_1, Z_2] \, \Phi + m (Z_1^2 + Z_2^2)   \right)
\, .
\end{align} 
It corresponds to giving a vacuum expectation value to $Z$'s via the mechanism of spontaneous symmetry breaking. Then, upon the shift by $\vev{Z}=O(m)$, the four-scalar interaction in \re{L} 
yields the first term in \re{mass} and the Yukawa coupling induces a mass term for the gauginos. The last term in the right-hand side of \re{mass} is required by supersymmetry 
\cite{Pestun:2007rz,Bobev:2013cja}. It breaks, however, the reflection positivity on the sphere \cite{Festuccia:2011ws}. 

The resulting action of the $\mathcal N =2^\ast$ model on the sphere takes the form 
\begin{align}\label{action}
S_{\mathcal N =2^\ast} = \int d^4 x\, \sqrt{\mbox{\sl g}}\, \left({\mathcal L}_0 +{\mathcal L}_{\rm int} +{\mathcal L}_{\rm mass} \right)\,,
\end{align}
where the first two terms inside the brackets are given by Eq.\ \re{L} with the fields redefined according to Eqs.\ \re{VectorMultipletFields} and \re{HyperFields}. 

To reveal the particle mass spectrum in this theory, we examine the action \re{action} at zero value of the gauge coupling. We get
\begin{align}
\label{LagrangianS4Nstar2}\notag
 S_{\mathcal N =2^\ast} \Big|_{g=0}
 = \int d^4 x\, \sqrt{\mbox{\sl g}}\,
 {\rm tr}\ 
\bigg[\frac12 F^{\mu\nu} F_{\mu\nu}- 2 i \lambda_{\alpha}^i {\not\!\!\mathcal{\nabla}} ^{\alpha\beta'}\bar\lambda_{\beta', i}+ \nabla^\mu \Phi \nabla_\mu \Phi + \frac{2}{R^2} \Phi^2
\\\notag
 +\nabla^\mu A_i  \nabla_\mu A_i +\nabla^\mu B_i \nabla_\mu B_i+ \lr{\mu_+^2 + \frac{2}{R^2}}A_i^2 + \lr{\mu_-^2 + \frac{2}{R^2}}B_i^2 
\\
 - 2 i \psi_{\alpha}^i {\slashed{\nabla}}_{\mu}^{\alpha\beta'}
\bar\psi_{\beta', i}
- i m \left( \psi^{i \alpha} \psi^i_\alpha + \bar\psi^{i \alpha'} \bar\psi^i_{\alpha'} \right)\bigg],
\end{align}
where the first and the following two lines encode the field content of the vector and hyper-multiplets, respectively. The kinetic terms for fields
involve the covariant derivatives on the sphere defined in \re{LBoperator} and \re{nabla}. Here we introduced two real (pseudo)scalar fields $A_i$ and $B_i$ 
\begin{align}
 Z_i= \frac{1}{\sqrt{2}} \left( A_i + i B_{i} \right)\,, \qquad (i=1,2)\,.
\end{align}
According to \re{HyperFields}, they originate from  the progenitor $\mathcal{N} = 4$ fields $(X_1, X_4)$ and $(X_2, X_5)$, respectively.
It follows from \re{LagrangianS4Nstar2} that $A_i$ and $B_i$   acquire complex ``masses', 
\begin{align}
\label{AandBmasses}
\mu^2_\pm = 
m^2 \pm i \frac{m}{R} 
\, ,
\end{align}
respectively. Different, though real, masses for (pseudo)scalars of the Wess-Zumino multiplet in the anti-de Sitter space were known to be enforced by supersymmetry
\cite{Burges:1985qq}.

By construction,  the  $\mathcal{N} = 2^\ast$ theory coincides with the $\mathcal N=4$ one for $m\to 0$. In the opposite limit, i.e.,  $m\to\infty$, the hypermultiplet becomes infinitely 
heavy and we recover the $\mathcal{N} = 2$ super Yang-Mills.

To simplify the analysis that follows, we set the radius of the four-sphere to be $R=1$. When needed, the dependence on $R$ can be easily restored by rescaling the mass and distances.

\subsection{Embedding coordinates}
\label{SectionEmbedding}

The action of the $\mathcal{N} = 2^\ast$ theory \re{action} is defined on the four-dimensional sphere. For the reason that will become clear in the next section, we have to rather generalize 
the definition \re{LagrangianS4Nstar2} to the $D-$dimensional sphere.

Instead of using local coordinates on the sphere (see Appendix~\ref{app:loc}), it is advantageous to embed $S^D$ in a flat Euclidean space  $\mathbb{R}^{D+1}$.  
Introducing the embedding coordinates $X_M$ (with $M=0,\dots,D$), the $D-$dimensional sphere of radius $R=1$ is defined as
\begin{align}\label{S^D}
S^D: \qqqquad
\sum_{M=0}^{D} X_M^2 = 1
\, .
\end{align}
Employing a stereographic projection
\begin{align}\label{stereo}
X_{0} = \frac{x^2 - 1}{x^2 + 1}  
\, , \qqqquad
X_\mu = \frac{2 x_\mu}{x^2 + 1}
\, , 
\end{align}
we can map the sphere to a hyperplane $X'_M=(0,x_\mu)$  with $\mu=1,\dots, D$ and $x^2 = x_\mu^2$. An important property of this transformation is that it is conformal in $\mathbb{R}^{D+1}$. 
Therefore, had the $\mathcal{N} = 2^\ast$ theory \re{LagrangianS4Nstar2} possessed conformal symmetry, we could apply \re{stereo} to obtain the same theory in the flat space-time $\mathbb{R}^{D}$. 

Since the conformal symmetry of the $\mathcal{N} = 2^\ast$ theory \re{action} is broken for $m\neq 0$, we are bound to study it on the sphere. Notice that at zero coupling, the Lagrangian of the vector 
multiplet, given by the first line in \re{LagrangianS4Nstar2}, does in fact enjoy the conformal symmetry. We exploit this fact in the next subsection to obtain a compact representation for free propagators of 
massless fields in the $\mathcal{N} = 2^\ast$ model.

A lot of activity has been devoted in the past to study field theories in the de Sitter space. This space can be obtained from \re{S^D} 
 by a Wick rotation, $X_{0} \to i X_{0}$,  
\begin{align}
{\rm dS}^D: \qqqquad \eta_{MN} X^M X^N = - 1\, , 
\end{align}
where $X^M$ are the embedding coordinates in $\mathbb{R}^{1,D}$ with the Minkowskian signature $\eta_{MN} = {\rm diag} (+,-,\dots,-)$. A convenient choice of local coordinates in the de Sitter space, 
widely used in applications to cosmology, looks as
\begin{align}
\label{EmbedToConfTime}
X^0  = - \frac{1 - z^2}{2 z^0}  \,, \qqqquad X^i  = \frac{z^i}{z^0}\,, \qqqquad  X^D = - \frac{1 + z^2}{2 z^0} \,,
\end{align}
where $i=1,\dots, D-1$ and $z^\mu$ (with $\mu = 0,1,\dots, D-1$) are the Minkowskian coordinates, $z^2 \equiv (z^0)^2 - (z^i)^2$.
The inverse transformation takes the form
\begin{align}
\label{EmbedToConfTimeInverse}
z^0 = - \frac{1}{X^0+X^D}
\, , \qqqquad
z^i = - \frac{X^i}{X^0+X^D}
\, .
\end{align}
We show in Appendices \ref{AppendixScalarPropagator} and \ref{AppendixFermionPropagator} that calculations on the sphere can sometimes be simplified by Wick rotating to de Sitter space and 
using the coordinates \re{EmbedToConfTime} instead.  
 
\subsection{Propagators of vector multiplet}

As a first step toward computing the circular Wilson loop in the $\mathcal N=2^\ast$ theory, we derive in the next two subsections expressions for free propagators of fields entering \re{LagrangianS4Nstar2}
 
We start with the massless sector of the theory. As was mentioned in the previous subsection, free propagators of massless fields on the sphere $S^D$ can be obtained from analogous expressions in flat space 
by a conformal transformation \re{stereo}.  For the scalar field $\Phi$ belonging to the vector multiplet we have \cite{Drummond:1975yc},
\begin{align}
\label{PhiProp}
D_\Phi (X_{12}) 
= \frac{\Gamma (\ft{D}2 -1)}{4 \pi^{D/2}} \frac{1}{|X_{12}|^{D-2}}
\, ,
\end{align}
with the notation $X_{12} = X_1 - X_2$ used throughout the paper.

Let us turn to the gauge propagator. The gauge field on the sphere is related by a coordinate transformation to its analogue in $\mathbb R^{D+1}$ 
\begin{align}\label{A-A}
A^{\mu}(x) = {\partial X_M \over \partial x_\mu}  A^M(X)\,,
\end{align}
where $x_\mu$ are the local coordinates on the sphere $S^{D}$ and $X_M$ are the embedding coordinates in $\mathbb R^{D+1}$. In the Feynman gauge, the propagator  $D_A^{MN} (X_{12}) 
=\vev{A^M(X_1) A^N(X_2)}$ obeys the same equation as the scalar propagator and, therefore, it reads
\cite{Adler:1972qq,Drummond:1977uy}
\begin{align}\label{D-gauge}
D_A^{MN} (X_{12}) = \eta^{MN} D_\Phi(X_{12}) 
\, ,
\end{align}
where $\eta^{MN} = {\rm diag} (+,\dots,+)$ is the metric tensor in $\mathbb R^{D+1}$.

To determine the massless gaugino propagators, it is convenient to combine two Weyl fermions into a Majorana one, $\Lambda=(\lambda_\alpha, \bar\lambda_{\alpha'})$, and  to work with 
four-components spinors and  Dirac matrices $\gamma^M$ (with $M=0,\dots,D$) in $\mathbb R^{D+1}$. As before, the fermion propagator on the sphere is related to that on $\mathbb R^{D+1}$ 
by~\cite{Adler:1972qq,Drummond:1977uy}
\begin{align}
\label{MasslessFermionProp}
D_\Lambda (X_1, X_2)
= \frac{\Gamma (\ft{D}2)}{2 \pi^{D/2}} \frac{U_1^{-1} {\not\!\!X}_{12} U_2}{|X_{12}|^{D}} 
\, ,
\end{align}
where the rotation matrices  $U_i= U (X_i)$ satisfy the following defining relations
\begin{align}
\label{GammaRotations}
U \gamma^a U^{-1} = e_{\mu}^{a}(x)\, {\partial \over \partial x_\mu} {\not\!\!X} 
\,  , \qqqquad
U \gamma^{D} U^{-1} = {\not\!\!X} \, ,
\end{align}
with the usual notation for ${\not\!\!X} \equiv \gamma^M X_M$, $x^\mu$ being the local coordinates on $S^D$ and the corresponding vielbeins $e_{\mu}^{a}(x)$ (with $a = 0,\dots, D-1$). The 
explicit form of the matrix $U(X)$ can be found in the Appendix \ref{AppendixFermionPropagator}, although is not important for our purposes.

\subsection{Propagators of massive hypermultiplet}

Moving on to the massive hypermultiplet, we can not rely on the conformal symmetry anymore. The details of calculations of the gaugino and scalar propagators entering the last two lines of 
\re{LagrangianS4Nstar2} can be found in the Appendices \ref{AppendixScalarPropagator} and \ref{AppendixFermionPropagator}, respectively. 

We start with the scalars  $A_i$ and $B_i$ (with $i=1,2$), which differ in their complex ``masses" \re{AandBmasses}.  Their propagators take the following form in the embedding coordinates
\begin{align}\label{D-AB}
D_A(X_{12}) = S_+(X_{12}^2) \,,\qqqquad D_B(X_{12}) = S_-(X_{12}^2) \,,
\end{align}
where a new notation is introduced for
\begin{align}
\label{PropAandBscalars}
S_\pm (X_{12}^2)
=
\frac{\Gamma (\ft{D}{2} - 1 \pm i m) \Gamma (\ft{D}{2} \mp i m)}{(4 \pi)^{D/2} \Gamma (\ft{D}{2})}
{_2 F_1}
\left.
\left(
{
\ft12 D - 1 \pm i m
\, , 
\ft12 D \mp i m
\atop
\ft12 D
}
\right| 1 - \frac{|X_{12}|^2}{4}
\right)
\, ,
\end{align}
to emphasize the relation between the two via the reflection transformation $m\to -m$. We verify that for $m\to 0$ the functions $S_\pm(X_{12}^2)$ coincide with the massless propagator \re{PhiProp}.

As in the massles case, it is convenient to introduce a Majorana fermion $\Psi=(\psi_\alpha,\bar\psi_{\alpha'})$. Referring to the Appendix \ref{AppendixFermionPropagator} for a thorough exposition of 
the calculations involved, we only quote here the result for the free propagator of the Majorana fermion of the massive hypermultiplet  
\begin{align}\label{ferm-prop}
D_\Psi (X_1, X_2)
=
U_1^{-1}
\bigg[
A (X_{12}^2) {\not\!\!\widetilde{X}}_{12}  {\not\!\!X}_2
+
B (X_{12}^2)
{\not\!\!X}_{12}
\bigg]
U_2
\, .
\end{align}
It involves the same rotation matrices as in the massless case \re{MasslessFermionProp} but contains an additional Dirac matrix structure, ${\not\!\!\widetilde{X}}_{12}  {\not\!\!X}_2$ with 
${\not\!\! \widetilde{X}}_{12} \equiv {\not\!\! {X}}_{1}+{\not\!\! {X}}_{2}$. The coefficient functions $A(X_{12}^2)$ and $B(X_{12}^2)$ are given in terms of $S_\pm$ introduced earlier, 
\begin{align}\notag
A (X_{12}^2)
&
=
\frac{
(\ft12 D - 1 + i m ) S_+ (X_{12}^2)
-
(\ft12 D - 1 - i m ) S_- (X_{12}^2)}{|X_{12}|^2 - 4} 
\, , \\[2mm]
\label{Bfermion}
B (X_{12}^2)
&
=
\frac{
(\ft12 D - 1 + i m ) S_+ (X_{12}^2)
+
(\ft12 D - 1 - i m ) S_- (X_{12}^2)}{|X_{12}|^2} 
\, .
\end{align}
We verify that for $m\to 0$ the function $A(X_{12}^2)$ vanishes and the expression for $D_\Psi (X_1, X_2)$ coincides with the massless propagator \re{MasslessFermionProp}. The 
properties of the functions $A (X_{12}^2)$ and $B (X_{12}^2)$ are discussed in the Appendix~\ref{app:MB}.  

\section{Circular Wilson loop}
\label{SectionWilsonLoop}

In this work, we study the circular supersymmetric  Wilson loop in $\mathcal N=2^\ast$ theory on the four-sphere. It is defined as~\cite{Maldacena:1998im}  
\begin{align}
W= {1\over N} \VEV{ \tr  P \exp \lr{i g \oint_C ds \left(\dot x_\mu(s) A^\mu(x(s))+i \Phi (x(s)) |\dot x(s)|\right)}} \,,
\end{align}
where the gauge field $A_\mu(x)$  and the real scalar $\Phi$ from the vector multiplet \re{VectorMultipletFields} are integrated along 
a great circle $C$ on the four-sphere of radius $R = 1$. Here $|\dot x(s)|=\lr{g^{\mu\nu} \dot x_\mu(s)\dot x_\nu(s)}^{1/2}$ with $\dot x_\mu(s)=\partial_s  x_\mu(s)$ and
$g^{\mu\nu}(x)$ being the metric tensor on the sphere.

Using \re{A-A}, we obtain an equivalent representation for the circular Wilson loop in the embedding coordinates
\begin{align}\label{DefWilsonLoop}
W= {1\over N} \VEV{ \tr  P \exp \lr{i g \oint_C ds \left(\dot X_M(s) A^M(X(s))+i \Phi (X(s)) |\dot X(s)|\right)}}\,,
\end{align} 
where $ |\dot X(s)|=(\dot X_M^2(s))^{1/2}$ and the great circle on the sphere $C$  can conveniently be parametrized as
\begin{align}
\label{Circle4D}
X_M (s) = (\cos s,\sin s,\bit{0})
\, ,
\end{align}
with an affine parameter $s$ ranging in the interval $0 \le s < 2\pi$. 

To lowest order in the coupling constant, we apply \re{PhiProp} and \re{D-gauge} for $D=4$ to get from \re{DefWilsonLoop}
\begin{align}\label{1-loop}
W
=
1 - \frac{g^2 C_F}{8 \pi^2} \int_0^{2 \pi} d s_1 \int_0^{2 \pi} d s_2 \frac{(\dot{X}_1  \cdot \dot{X}_2) - |\dot{X}_1| |\dot{X}_2|}{|X_1 - X_2|^2} + O(g^4)
= 
1 +  \frac{1}{4}g^2 C_F + O(g^4)
\, ,
\end{align}
where $X_i \equiv X(s_i)$ and $C_F = (N^2-1)/(2N)$ is the quadratic Casimir in the fundamental representation of the $SU(N)$. Here the first term in the numerator of the integrand comes from the 
gauge field and the second one from the scalar. 

The one-loop correction to \re{1-loop} is not sensitive to the hypermultiplet and, therefore, it is the same both in the $\mathcal N=2^\ast$ and $\mathcal N=4$ theories. The massive hypermultiplet \re{HyperFields} 
contributes to \re{DefWilsonLoop} starting at order $O(g^4)$ and, as a consequence, the circular Wilson loops in the two theories differ already at two loops. In the $\mathcal N=4$ theory, the circular Wilson loop 
can be found exactly at arbitrary value of the gauge coupling $g^2$ and finite $N$. At weak coupling, it was demonstrated by an explicit calculation in \cite{Erickson:2000af} (see also 
\cite{Drukker:2000rr,Plefka:2001bu,Arutyunov:2001hs}) that $W$ receives a nonzero contribution from ladder diagrams only and it can be expressed as a matrix integral. In the planar limit,  the planar 
circular Wilson loop reads
\begin{align}
\label{N4WLexact}
W_{\mathcal{N} = 4} 
\stackrel{N \to \infty}{=}   \frac{2}{\sqrt{g^2 N}} I_1 (\sqrt{g^2 N}) = 1 + {\lambda\over 8} + {\lambda^2\over 192} + O(\lambda^3)
\, ,
\end{align}
where $I_1$ is the modified Bessel function and $\lambda=g^2 N$ is the 't Hooft coupling.

The relation \re{N4WLexact} was proved in Ref.~\cite{Pestun:2007rz} using supersymmetric localization. It was also generalized there to the case of the massless superconformal $\mathcal{N} = 2$ 
Yang-Mills theory with fundamental hypermultiplets and the $\mathcal{N} = 2^\ast$ theory on the sphere. The former result was tested in Ref.\  \cite{Andree:2010na,Billo:2019fbi} against explicit three-loop calculation 
of the Wilson loop in flat space.  However, up to now not a single study was dedicated to testing the localization predictions in the $\mathcal{N} = 2^\ast$ theory. The reason for this is quite simple: in virtue of 
conformal symmetry, all of perturbative checks without exceptions were performed in flat space. In $\mathcal{N} = 2^\ast$ theory this simplification does not hold and one has to perform fullfledged
calculations on the sphere with massive propagators. As we discussed in the previous section, the latter are quite involved on their own. 

\subsection{Prediction from localization}

The circular Wilson loop in the $\mathcal{N} = 2^\ast$ theory is a finite function of the gauge coupling and the dimensionless parameter $mR$. Although we put $R=1$ for simplicity, the dependence on $R$ 
can be easily restored by replacing $m\to m R$. The conformal symmetry of $\mathcal{N} = 2^\ast$ theory is explicitly broken by the mass deformation. As a consequence, in distinction to the $\mathcal N=4$  
theory, one expects that the circular Wilson loop should admit different forms in the $\mathcal{N} = 2^\ast$ case on the sphere and in the flat space. We will show this explicitly below.  

As was alluded to above, the circular Wilson loops are identical in the $\mathcal N=4$ and $\mathcal N=2^\ast$ theories at one loop. The difference emerges only at two loops due to massive hypermultiplets 
circulating in virtual loops. Following Refs.~\cite{Andree:2010na,Billo:2017glv}, we find it convenient to consider the difference of the Wilson loops in the two models, $W_{\mathcal N=2^\star} - W_{\mathcal N=4}$. 
This allows us to avoid computation of diagrams with the massless $\mathcal{N} = 2$ vector multiplet in internal subgraphs and, thus, reduce significantly the number of contributing Feynman integrals.
 
The supersymmetric localization  predicts the following result for the difference of the circular Wilson loops at weak coupling   \cite{Pestun:2007rz}
\begin{align}\label{diff}
\Delta W_{S^4} \equiv
W _{\mathcal N=2^\star} - W _{\mathcal N=4}  = {g^4 C_F N\over 16\pi^2} f(m) + O(g^6)\,,
\end{align}
where we inserted the subscript $S^4$ to indicate that the gauge theories are defined on the four-sphere. Writing down the color factor in the right-hand side of \re{diff} in terms of the quadratic Casimir 
$C_F=(N^2-1)/(2N)$, we made use of the known property of non-Abelian exponentiation for Wilson loops in gauge theories~\cite{Gatheral:1983cz,Frenkel:1984pz}. 

The function $f(m)$ carries the dependence on the mass parameter.~\footnote{For the four-sphere of arbitrary radius $R \neq 1$, this function obviously depends on the dimensionless parameter $m R$.}
 Since it is independent of the rank of the gauge group $N$, it can be derived from the localization formulas for the $SU(2)$ gauge group. This gives
\begin{align}
\label{LocalizationFiniteM}
f(m) = \frac12\left[ \psi(1-i m)+\psi(1+ i m)-i m\, \psi'(1-i m)+i m\, \psi'(1+i m)+2 \gamma \right]\,,
\end{align}
where $\psi(x) = (\log \Gamma (x))'$ is the digamma function and $\gamma$ is the Euler-Mascheroni constant.  For future reference, we present its expansion in two different asymptotic regions:
\begin{align} \label{loc-small}
&  f=3 m^2 \zeta (3)-5 m^4 \zeta (5)+O\left(m^6\right)\,, && \text{for $m \ll 1$,}
\\[2mm]\label{loc-large}
&  f=  \log m +  \gamma + 1 -\frac{1}{12 m^2}-\frac{1}{40 m^4} + O\left(\frac{1}{m^6}\right)\,,  && \text{for $m \gg 1$.} 
\end{align}

The logarithm in the last relation is related to the beta-function in pure $\mathcal N=2$ super Yang-Mills theory.  Indeed, for $m\to\infty$, the hypermultiplet becomes infinitely heavy and 
$W_{\mathcal N=2^\star}$ should match the Wilson loop in the pure $\mathcal N=2$ super Yang-Mills theory endowed with an ultraviolet cut-off $m$. The coupling constant in the latter 
theory depends on $m$ and this dependence is driven by the known one-loop-exact beta-function.  The logarithmically enhanced term in the expression for
\begin{align}
W_{\mathcal N=2} = 1 +  \frac{1}{4}g^2 C_F + { g^4 C_F N\over 16\pi^2} (\log m + O(m^0) ) +O(g^6)
\end{align}
has to be proportional to the beta function in order for $W _{\mathcal N=2}$ to remain finite as $m\to\infty$.

\begin{figure}[t!]
\begin{center}
\mbox{
\begin{picture}(0,130)(225,0)
\put(0,-420){\insertfig{26}{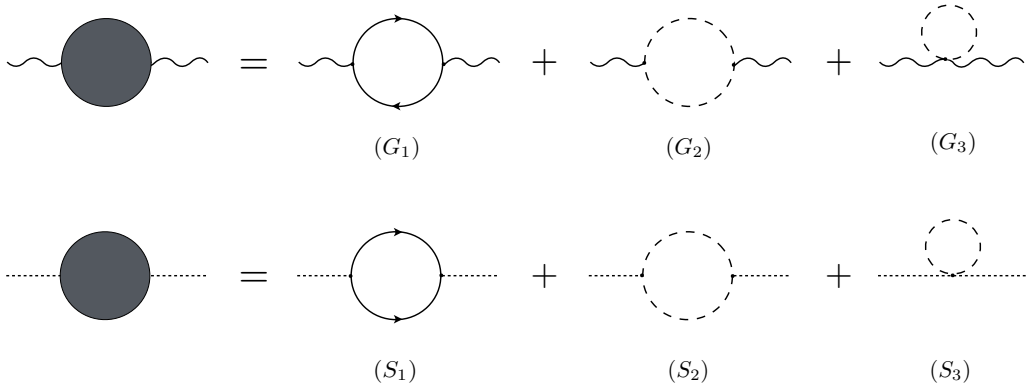}}
\end{picture}
}
\end{center}
\caption{\label{PolOoperatorsFig} 
One-loop corrections to the gauge field and scalar propagators. The wavy and dotted lines denote, respectively, the gluons and massless scalars belonging to the vector multiplet. 
The solid and  dashed lines represent the massive fermions and scalars from the matter hypermultiplet.}
\end{figure}
 
Our goal in this work is to verify \re{diff} by computing $\Delta W_{S^4}$ to order $O (g^4)$. As we just explained, the difference comes from graphs with the massive hypermultiplet circulating in 
loops. To two loop order, the only graphs we have to account for are those describing one-loop correction to propagators of the gauge field and massless scalar belonging to the vector multiplet,
$ D_m^{(1),MN}$ and $ D_m^{(1)}$, respectively (see Fig.\ \ref{PolOoperatorsFig}). Their contribution to the difference of the Wilson loops \re{diff} is 
\begin{align}\notag\label{W-W}
& \Delta W_{S^4}=  -
\frac12{g^2 C_F} \int_0^{2 \pi} d s_1 \int_0^{2 \pi} d s_2
\\[2mm]
& \qquad \times \left[\dot X_{1,M} \dot X_{2,N} D_m^{(1),MN}(X_1,X_2) -  |\dot{X}_1| |\dot{X}_2| D_m^{(1)}(X_1,X_2)  \right]- (m=0)\,,
\end{align}
where $X_{i,M} \equiv X_M(s_i)$ is given by \re{Circle4D}. Here the last term in the right-hand side subtracts the contribution at $m=0$, so that this difference vanishes for $m\to 0$.

Each individual graph shown in Fig.\ \ref{PolOoperatorsFig} depends on the mass parameter $m$ and develops an ultraviolet divergence. The divergences cancel, however, in the right-hand side 
of Eq.\ \re{W-W}, after we subtract the contribution of the same graphs with $m=0$. Still, to define the contribution of individual diagrams, we have to introduce a regularization. For this purpose we 
employ conventional dimensional regularization and perform all calculations on the sphere $S^D$ with $D=4-2\epsilon$.

\subsection{Circular Wilson loop in flat space}
\label{FlatSpaceSection}

In this subsection, we compute the difference $W _{\mathcal N=2^\star} - W _{\mathcal N=4}$  in flat space and compare it with the analogous expression on the sphere predicted by the 
localization \re{diff}. As was explained at the beginning of this section, we anticipate the two expressions to be different due to conformal symmetry breaking in the $\mathcal N=2^\ast$ theory.

To lowest order in gauge coupling, the difference $\Delta W_{\mathbb R^4}$ is given by the same relation \re{W-W} with $X_M=(x_\mu, 0)$ and $x_\mu$ being local coordinates in $\mathbb R^4$.
The most concise and efficient way to perform calculations in flat space is to take advantage of translational invariance and pass to the reciprocal momentum space via the Fourier transform
\begin{align} 
& D_m^{(1),\mu\nu}(x_{12})= g^2 N\int {d^D p\over (2\pi)^D}\e^{i p \cdot x_{12}} {\Pi_G^{\mu\nu}(p)\over (p^2)^2}  \,,\qqqquad
\end{align}
where we factored out the dependence on the gauge coupling and the color factor. 
A similar relation holds for the scalar propagator $D_m^{(1)}(x_{12})$. Thus, there are two polarization operators we have to evaluate, one for the 
gauge field, $\Pi_G^{\mu\nu}(p)$,  and another for the massless scalar, $\Pi_\Phi(p)$. Let us consider them in turn.  

The Feynman diagrams contributing to the gluon polarization operator at one loop are depicted in the top row of Fig.\ \ref{PolOoperatorsFig}. In dimensional regularization with $D=4-2\epsilon$, the 
contribution from the graphs in $G_1$, $G_2$ and $G_3$ reads, respectively, in the momentum representation  
\begin{align}
\Pi_G^{\mu\nu} (p) 
= 
\int \frac{d^D k}{(2 \pi)^D}
\left[ 
- \frac{\tr [(\not\!k +i m) \gamma^\mu (\not\!k + \not\!p+ i m) \gamma^\nu]}{(k^2 + m^2) [(k + p)^2 + m^2]} 
+ \frac{ 2(2k + p)^\mu  (2k + p)^\nu}{(k^2 + m^2) [(k + p)^2 + m^2]}
-   \frac{4 g^{\mu\nu} }{k^2 + m^2}
\right].
\end{align}
The relative multiplicative factors are as follows: $(-1)$ in front of the first term comes from the Dirac-Fermi statistics, the factor of 2 in front of the second term is from two pairs of real (pseudo)scalars 
$A_i/B_i$ in the hypermultiplet, and, finally, 4 is from the same two pairs of scalars inside the tadpole. Due to the gauge invariance, the polarization operator is transverse, 
\begin{align}
\Pi_G^{\mu\nu} (p) = \left( g^{\mu\nu} - \frac{p^\mu p^\nu}{p^2}\right)  \pi_m (p^2)
\, .
\end{align}
Contracting both sides with $g_{\mu\nu}$, we find the integral representation for $\Pi_m (p^2)$,
\begin{align}
\label{PiOneLoop}
\pi_m (p^2) = \int \frac{d^D k}{(2 \pi)^D} \frac{2 p^2}{(k^2 + m^2) [(k + p)^2 + m^2]}
\, .
\end{align}
Integration over the large loop momentum yields a pole $1/(D-4)$ but it cancels in the difference $\pi_m (p^2)-\pi_0 (p^2)$.
 
The polarization operator for the massless scalar receives contributions from the Feynman diagrams in the second row of Fig.\ \ref{PolOoperatorsFig}. It reads
\begin{align}
\label{FlatSpaceYukawa}
\Pi_\Phi (p^2)
&
=  \int \frac{d^D k}{(2 \pi)^D}\left[  \frac{4(k \cdot (k + p) - m^2)}{(k^2 + m^2)[(k + p)^2 + m^2]}
+ \frac{8m^2}{(k^2 + m^2)[(k + p)^2 + m^2]} - \frac{4}{k^2 + m^2}\right]
\, ,  
\end{align}
for the Yukawa coupling ($S_1$), three-scalar interaction ($S_2$) and tadpole ($S_3$), respectively. Here the multiplicities of the hypermultiplet fields get
altered by the strengths of couplings as governed by the Lagrangian \re{action} in the limit $R\to\infty$. Adding up all contributions, we find
\begin{align}
\Pi_\Phi (p^2) = \pi_m (p^2)
\, ,
\end{align}
with $\pi_m (p^2)$ given in Eq.\ \re{PiOneLoop}, in agreement with expectations based on supersymmetry.

Combining together the above relations, we obtain from \re{W-W}
\begin{align}\label{W-R4}
\Delta W_{\mathbb R^4}
 = -
\frac12{g^4 C_F N} \int_0^{2 \pi} d s_1 \int_0^{2 \pi} d s_2\,\left[(\dot x_{1} \cdot \dot x_{2})  -  |\dot{x}_1| |\dot{x}_2|  \right] D^{(1)}(x_{12}^2)\,,
\end{align}
where the notation was introduced for 
\begin{align}\notag\label{D1}
D^{(1)}(x_{12}^2)
& =\int {d^4 p\over (2\pi)^4}\e^{i p x_{12}} { \pi_m (p^2)- \pi_0 (p^2)\over (p^2)^2}
\\
&=\int {d^4 p\over (2\pi)^4}{\e^{i p x_{12}}\over p^2} \int_0^1 {d\xi \over 8\pi^2}   \left[ \frac{\Gamma (2 - D/2)}{(\xi(1-\xi) p^2 + m^2)^{2 - D/2}} - (m=0)\right].
\end{align}
Here in the second relation, we replaced $\pi_m(p^2)$ with its integral representation \re{PiOneLoop} and performed the loop momentum integration
using Feynman parametrization. To proceed further, we use the Mellin-Barnes representation for the first term inside the brackets in \re{D1}
\begin{align}
\label{MBdenominator}
(X + Y)^{- \nu} = \frac{1}{\Gamma (\nu)} \int \frac{d j}{2 \pi i} \Gamma (- j) \Gamma (j + \nu) \frac{Y^{j}}{X^{j + \nu}}
\, ,
\end{align}
where the integration contour runs parallel to the imaginary axis and separates poles generated by the two $\Gamma-$functions in the integrand.

Substituting \re{D1} into \re{W-R4} and performing sequentially the integrations over the momentum and parameters $s_1,s_2$, we finally obtain 
\begin{align}\label{diff-flat}
\Delta W_{\mathbb R^4} = - {g^4 C_F N\over 64\pi^2}\int_{-\delta-i\infty}^{-\delta+i\infty}{dj\over 2\pi i}
(2m)^{2 j+2} \frac{ \Gamma^3 (-j-1) \Gamma \left(j+\frac{3}{2}\right)}{\Gamma \left(\frac{1}{2}-j\right) \Gamma (j+1)}\,,
\end{align}
where 
${\rm Re} \,j=-\delta$ and $0<\delta< 1$. We can use this relation to find asymptotic behavior of 
$\Delta W_{\mathbb R^4}$ at small and large values of $m$.
At large $m$, we move the integration contour to the left and pick up residues at poles $j=-1,-3/2,-5/2, \dots$
\begin{align}\label{as1}
\Delta W_{\mathbb R^4} \stackrel{m\gg 1}{=} {g^4 C_F N\over 16\pi^2}\left[\log \bar m +1 +\frac{\pi }{16 m}+\frac{3 \pi }{2048
   m^3} + O\lr{\frac{1}{m^5}} \right],
\end{align}
where  $\bar m = m  \e^{\gamma}/2$.
At small $m$, we move the integration contour to the right and observe that the integrand has third order poles at $j=0,1,\dots$. Evaluating the residue at these poles, we obtain
\begin{align}\notag\label{as2}
\Delta W_{\mathbb R^4} \stackrel{m\to 0}{=} {g^4 C_F N\over 16\pi^2}\Big[ &\ m^2 \left(\log ^2 \bar m -\log \bar m +\frac{\pi ^2}{12}\right)
\\
+ {}&\,  m^4 \left(\frac{3 \log ^2 \bar m}{8}-\frac{5 \log \bar m}{16}+\frac{\pi
   ^2}{32}-\frac{9}{32}\right)+ O(m^6)\Big].
\end{align}
This expression vanishes for $m\to 0$ since in this limit $\mathcal N=2^\ast$ coincides with the $\mathcal N=4$ theory.

The relations \re{as1} and \re{as2} are in agreement with results obtained in Refs.~\cite{unpublished1,unpublished2}. 

Let us compare $\Delta W_{\mathbb R^4}$ with corresponding expression
for the difference of the Wilson loops on the sphere found via the localization, see Eqs.~\re{diff} -- \re{loc-large}.
As explained above, at large $m$, the leading $O(\log m)$ term in $\Delta W_{S^4}$ and  $\Delta W_{\mathbb R^4}$ is driven by the beta-function and, therefore, it should be the same. The 
two expressions differ starting from $O(m^0)$ terms. It is interesting to note that subleading, i.e., power suppressed, corrections to the circular Wilson loop have a form different on the sphere than 
in the flat space:  for $\Delta W_{S^4}$ and $\Delta W_{\mathbb R^4}$ they run, respectively, in even and odd powers of $1/m$. The difference also persists at small $m$, though expansions of 
both $\Delta W_{S^4}$ and $\Delta W_{\mathbb R^4}$ run in powers of $m^2$, in the latter case the accompanying coefficients are enhanced by powers of $\log m$.

The underlying reason why the circular Wilson loop is different in the flat space and on the sphere is due to conformal symmetry breaking in $\mathcal N=2^\ast$ SYM for $m\neq 0$. Naively one 
may expect that $\Delta W_{S^4}$ and $\Delta W_{\mathbb R^4}$ should coincide in the limit when the radius of the four-sphere becomes large, or equivalently $m R\gg 1$. We just demonstrated 
that this is not the case. To see why this happens, consider the massive scalar propagator on the sphere \re{PropAandBscalars}. Restoring the dependence on the radius $R$, we observe 
that the propagator is a function of two dimensionless parameters $mR$ and $X_{12}^2/R^2$, with $X_{12}^2$ being a chordal distance on the sphere. Going to the limit $R\to\infty$ with $m$ held 
fixed, we have to distinguish two cases: $X_{12}^2/R^2\to 0$ and $X_{12}^2/R^2=O(R^0)$ corresponding to the short and large distances on the sphere. The scalar propagator coincides with its 
flat-space counterpart in the former case only. The latter situation is realized for the circular Wilson loop when the scalar propagator is stretched across the great circle of the sphere.

\section{Circular Wilson loop at two loops}
\label{SectionMB}

In this section, we apply the relation \re{W-W} and work out the representation for $ \Delta W_{S^4}$ to order $O(g^4)$  in terms of Feynman integrals on the sphere. The technique for computing 
these integrals is addressed in the following sections. 

Applying \re{W-W}, we have to compute one-loop corrections to propagators of the gluon and massless scalar. The corresponding Feynman diagrams are shown in Fig.~\ref{PolOoperatorsFig}. In 
close analogy with the flat-space analysis of Section \ref{FlatSpaceSection}, their contribution is given by the product of propagators integrated over the position of the interaction vertices on the 
sphere. Performing calculations, we will use the embedding coordinates, which prove to be very efficient.

In what follows we use the notation $X_N(s)$ (with $0\le s< 2\pi$) for points on the great circle of the sphere and $Z_{i,N}$ for the coordinates of the integration vertices $\sigma_i$.
The integration measure on $S^D$ is
\begin{align}\label{measure}
\int d\sigma_i \equiv \int d^{D+1}  Z_i \,\delta(1-|Z_i|^2)\,,
\end{align}
where $|Z_i|^2= \eta^{NM} Z_{i,N} Z_{i,M}$ and $\eta^{NM} = {\rm diag}\,(+,\dots,+)$. The metric on the sphere is conformally flat,
\begin{align}
ds^2 = \eta^{NM} d Z_N d Z_M =  g^{\mu\nu}(x) dx_\mu dx_\nu\,,
\end{align}
where $x_\mu$ are some local coordinates on $S^D$ and $g^{\mu\nu}(x) = \eta^{NM} (\partial Z_M/\partial {x_\mu})  (\partial Z_N/\partial {x_\nu})$ is the corresponding metric tensor. We will also 
need the following tensor
\begin{align}\label{Q-def}
Q_{NM}(Z) = g_{\mu\nu}(x) {\partial Z_M\over \partial {x_\mu}}  {\partial Z_N\over\partial {x_\nu}} = \eta^{NM} - Z^N Z^M\,,
\end{align}
that obeys the relations $Q_{NM} Z^M=0$ and $Q_{NM_1} \eta^{M_1M_2}Q_{M_2M}=Q_{NM}$ and serves as a projector onto a hyperplane orthogonal to $Z^M$.

\subsection{Corrections to gluon propagator}

Using the Feynman rules previously presented in Section \ref{sect:N=2}, we deduce the following representation for the contribution of diagrams displayed in the first line of  Fig.\ \ref{PolOoperatorsFig}
\begin{align}\label{Int1}
\dot X_{1,M} \dot X_{2,N} D_m^{(1),MN}(X_1,X_2) = g^2 N\int d \sigma_1 d \sigma_2 D_\Phi (X_1 - Z_1) D_\Phi (X_2 - Z_2) \Pi_G (Z_1, Z_2)\,,
\end{align}
where we replaced the gauge propagator with \re{D-gauge} and introduced a notation for the one-loop polarization operator contracted with the tangent vectors
$\dot X_1$ and $\dot X_2$. It can be split into the sum of three terms
\begin{align}
\Pi_G = \Pi_{G_1} + \Pi_{G_2} + \Pi_{G_3}
\, ,
\end{align}
according to the graphs shown in the top row of Fig.\ \ref{PolOoperatorsFig}, respectively.

We start with the contribution of the fermion loop. It takes the form
\begin{align}\label{Pi-G1}
\Pi_{G_1} = -(\dot{X}_1\cdot \partial_{x_1}^\mu Z_1 ) 
(\dot X_2 \cdot \partial_{x_2}^\nu Z_2) e_{a,\mu}(x_1) e_{b,\nu}(x_2) \tr\left[\gamma^a D_\Psi(Z_1,Z_2) \gamma^b  D_\Psi(Z_2,Z_1) \right],
\end{align}
where the first two factors come from the contraction of the gauge field \re{A-A} with the tangent vectors and $e_{a,\mu}(x)$ are the vielbeins. 
Replacing the fermion propagator $D_\Psi$ with its explicit expression \re{ferm-prop}, we encounter the following expression
\begin{align}\notag
(\dot{X}_1\cdot \partial_{x_1}^\mu Z_1)e_{a,\mu}(x_1) U_1 \gamma^a U_1^{-1}
& = (\dot{X}_1\cdot \partial_{x_1}^\mu Z_1)e_{a,\mu}(x_1) e_{\nu}^{a}(x_1)\, {\partial_{x_1}^\nu Z_{1,M}} \gamma^M
\\[2mm]
& = (\dot{X}_1\cdot \partial_{x_1}^\mu Z_1)g_{\mu\nu}(x_1) {\partial_{x_1}^\nu Z_{1,M}} \gamma^M = \dot{X}_1^N Q_{NM}(Z_1) \gamma^M\,.
\end{align}
Here, in the first line, we applied \re{GammaRotations}, while in the second line, we used \re{g-co} and \re{Q-def}. Taking into account the last relation, we get from 
\re{Pi-G1}
\begin{align}\label{Pi-G1-int}
\Pi_{G_1} = - \dot{X}_1^{N_1} Q_{N_1M_1}(Z_1) \dot X_2^{N_2} Q_{N_2M_2}(Z_2)  \tr\left[\gamma^{M_1} \widehat D_\Psi(Z_1,Z_2) \gamma^{M_2}\widehat D_\Psi(Z_2,Z_1) \right],
\end{align}
where $\widehat D_\Psi(Z_1,Z_2)$ is the fermion propagator \re{ferm-prop} stripped down from the rotation matrices $U$ and $U^{-1}$
\begin{align}
 \widehat D_\Psi(Z_1,Z_2) = 
A (Z_{12}^2) {\not\!\!\widetilde{Z}}_{12}  {\not\!\!Z}_2
+
B (Z_{12}^2)
{\not\!\!Z}_{12}\,.
\end{align}
The relation \re{Pi-G1-int} has a form that is similar to the fermion loop correction to the gauge propagator in a flat space. 
One important difference, however, is that all vector indices are contracted using the tensor $Q_{NM}$. This ensures gauge invariance of $\Pi_{G_1}$ \cite{Adler:1972qq,Drummond:1977uy}.  

Going through calculations of \re{Pi-G1-int}, we obtain 
\footnote{Arriving at this relation, we neglected terms containing total derivatives with respect to $s_1$ and $s_2$. They yield vanishing contributions
in the circular Wilson loop upon the integration over the contour in Eq.\ \re{W-W}.}
\begin{align}
\label{PiG1}
\Pi_{G_1} (Z_1, Z_2)
=&
-16 (\dot{X}_1 \cdot \dot{X}_2) A^2 (Z_{12}^2) 
\nonumber\\[2mm]
&+4 [ Z_{12}^2 (\dot{X}_1 \cdot \dot{X}_2)+2  (\dot{X}_1 \cdot Z_2)(\dot{X}_2 \cdot Z_1)] [A^2 (Z_{12}^2) +B^2 (Z_{12}^2)] \,,
\end{align}
where the functions $A(Z^2)$ and $B(Z^2)$ are given by \re{Bfermion}.

The calculation of the scalar loop goes along the same lines
\begin{align}\notag
\Pi_{G_2} 
& = 2\dot{X}_1^{N_1} Q_{N_1M_1}(Z_1) \dot X_2^{N_2} Q_{N_2M_2}(Z_2) 
\\[2mm]
& \times
\sum_{\sigma=\pm}
\left[
\partial_{Z_1}^{M_1} \partial_{Z_2}^{M_2} S_\sigma(Z_{12}^2) S_\sigma(Z_{12}^2)  
-\partial_{Z_1}^{M_1} S_\sigma(Z_{12}^2) \partial_{Z_2}^{M_2} S_\sigma(Z_{12}^2)  
\right].
\end{align}
As in the previous case, this expression is similar to its analogues in the flat space. Here the two terms in the sum describe the contribution of the (pseudo)scalars $A_i$ and $B_i$, respectively. 
Their propagators are defined in \re{D-AB} and \re{PropAandBscalars}. The functions $S_\pm$ satisfy nontrivial relations (see Eqs.\ \re{SfromAB} and \re{difS}) and they can be expressed in terms 
of the functions $A(Z^2)$ and $B(Z^2)$ introduced in Eq.~\re{Bfermion}. Using these results, we can cast $\Pi_{G_2}$ into the form
\begin{align}
\label{PiG2}
\Pi_{G_2} (Z_1, Z_2)
&= -8 (\dot{X}_1 \cdot Z_2) (\dot{X}_2 \cdot Z_1) [A^2(Z_{12}^2) + B^2(Z_{12}^2)]\,.
\end{align}

Combining together \re{PiG1} and \re{PiG2}, we find that the sum $\Pi_{G_1}+\Pi_{G_2}$ takes a remarkably simple form
\begin{align}\label{g1+g2}
\Pi_{G_1+G_2} (Z_1, Z_2)  &= -4(\dot{X}_1 \cdot \dot{X}_2)  \left[\widetilde Z_{12}^2A^2 (Z_{12}^2) -   Z_{12}^2B^2 (Z_{12}^2)\right]\,,
\end{align}
where $\widetilde Z_{12}^2=(Z_1+Z_2)^2=4-Z_{12}^2$ and  $Z_{12}=Z_1-Z_2$.

Finally, the tadpole contribution to the gauge propagator is localized at $Z_1=Z_2$ and it is proportional to propagators of the massive scalars at zero separation
\begin{align}\label{tad1}
\Pi_{G_3} (Z_1, Z_2)
&=  -2 [S_+ (0) + S_- (0)]  (\dot{X}_1 \cdot \dot{X}_2) \delta (Z_{12})\,,
\end{align}
where the delta function on the sphere is defined as $\int d \sigma_1 \delta (Z_{12}) f(Z_1) = f(Z_2)$ for an arbitrary test function.

\subsection{Corrections to scalar propagator}

Analogously, for the one-loop correction to the massless scalar propagator, we get 
\begin{align}\label{Int2}
D_m^{(1)}(X_1,X_2) = g^2 N \int d \sigma_1 d \sigma_2 D_\Phi (X_1 - Z_1) D_\Phi (X_2 - Z_2) \Pi_S (Z_1, Z_2)\,,
\end{align}
where the scalar polarization operator is again a sum of three contributions
\begin{align}
\Pi_S = \Pi_{S_1} + \Pi_{S_2} + \Pi_{S_3}
\, ,
\end{align}
for the Yukawa, triple scalar and tadpole contributions shown in the bottom row of Fig.\ \ref{PolOoperatorsFig}, respectively.
They read 
\begin{align}\notag\label{tad2}
\Pi_{S_1} (Z_1, Z_2)
&= -4\left[\widetilde Z_{12}^2 A^2 (Z_{12}^2) + Z_{12}^2 B^2 (Z_{12}^2) \right]
\, , 
\\[.8mm] \notag
\Pi_{S_2} (Z_1, Z_2)
&=-4 m^2 [S^2_+ (Z_{12}^2) + S^2_- (Z_{12}^2)] 
\, , \\[2mm]
\Pi_{S_3} (Z_1, Z_2)
&=  - 2 [S_+ (0) + S_- (0)]  \delta (Z_{12})
\, .
\end{align}
The first expression should be compared with the analogous equation in \re{g1+g2}.

\subsection{General expression}

Combining together Eqs.\ \re{Int1} and \re{Int2}, we conclude that the one-loop correction to the linear combination of gauge and scalar propagators in \re{W-W} is given by
\begin{align}\label{sum-D}
\dot X_{1,M} \dot X_{2,N} D_m^{(1),MN}(X_1,X_2) -  |\dot{X}_1| |\dot{X}_2| D_m^{(1)}(X_1,X_2)  = g^2 N \left[  D_{\rm loop}  + D_{\rm tadpole} \right]\,,
\end{align}
where $D_{\rm loop}$ and $D_{\rm tadpole}$ describe contributions from diagrams in Fig.\ \ref{PolOoperatorsFig} containing loops and tadpoles, respectively. 

We take into account \re{tad1} and \re{tad2} to find
\begin{align}\notag
& D_{\rm tadpole}= -2 [S_+ (0) + S_- (0)] ((\dot X_1 \cdot \dot X_2)-|\dot X_1 ||\dot X_2|) \int d \sigma_1   D_\Phi (X_1 - Z_1) D_\Phi (X_2 - Z_1)\,,
\\
 \label{D-loop}
& D_{\rm loop} = -4\int d \sigma_1 d \sigma_2 D_\Phi (X_1 - Z_1) D_\Phi (X_2 - Z_2)
\left[(\dot X_1 \cdot \dot X_2) \Pi_1(Z_{12}^2) + |\dot X_1 ||\dot X_2| \Pi_2(Z_{12}^2)  
\right].
\end{align}
Here in the first relation, we used $\delta(Z_{12})$ to integrate over $Z_2$ and, in the second relation, we introduced notations for 
\begin{align}\notag\label{Pi's}
& \Pi_1 =\widetilde Z_{12}^2A^2 (Z_{12}^2) -   Z_{12}^2B^2 (Z_{12}^2)\,,
\\[2mm]
& \Pi_2 =\widetilde Z_{12}^2A^2 (Z_{12}^2) +  Z_{12}^2B^2 (Z_{12}^2) + m^2 [S^2_+ (Z_{12}^2) + S^2_- (Z_{12}^2)] \,,
\end{align}
where $\widetilde Z_{12}^2=4-Z_{12}^2$.
These equations involve functions defined in \re{PropAandBscalars} and \re{Bfermion}.

In the next section, we describe a technique for computing integrals entering \re{D-loop}.  

\subsection{Propagators at zero separation}

Before we turn to computing all integrals, let us examine the contribution from the tadpoles \re{D-loop}. It depends on the scalar propagators $S_\pm(0)$ at coincident points. 
To find them, it is convenient to use the relation \re{Bfermion} for $X_{12}^2\to 0$ (see also \re{SfromAB})
\begin{align}
\label{SfromABatZero}
S_+(0) =- {4A(0) \over D-2+2im}\,,\qqqquad 
S_-(0) = {4A(0)\over D-2-2im}
\, , 
\end{align}
so that the sum of the two propagators becomes
\begin{align}\label{sum-S}
S_+(0)+S_-(0) = \frac{16 i m A(0)}{(D-2)^2+4 m^2}
\, .
\end{align}
To find $A(0)$, it is instructive to use the Mellin-Barnes representation of $A(X_{12}^2)$ derived in the Appendix~\ref{app:MB}. As a result, we find (see Eq.~\re{app-A0})
\begin{align}\label{A0}
A(0) =- {i m\over 2(4\pi)^{D/2}}  {\Gamma(1- \ft12 D)\Gamma(\ft12 D+im)\Gamma(\ft12 D-im)\over \Gamma(1+im)\Gamma(1-im)} 
\, .
\end{align}
For $D=4-2\epsilon$, we expand \re{sum-S} in powers of $\epsilon$ to get
\begin{align}\label{S-tp}
S_{0}\equiv -2\lr{S_+(0)+S_-(0)} = {m^2\over 4\pi^2} \lr{4\pi \e^{\gamma}}^\epsilon \left[ {1\over \epsilon} + 
   \left(1-H_{i m}-H_{-i m}\right)+O(\epsilon)\right] ,
\end{align}
where $H(x)=\psi(x+1) + \gamma$ are the harmonic numbers. In the above equation, the pole $1/\epsilon$ has a UV origin. In the two-loop expression for the circular Wilson loop, it  cancels 
against contributions of other diagrams in Fig.~\ref{PolOoperatorsFig}.

\section{Comparison with localization}\label{sect:comp}

In this section, we compute the integrals \re{D-loop} on the sphere and, then, use them to evaluate the circular Wilson loop \re{W-W}. 

Echoeing \re{sum-D}, we can split $\Delta W_{S^4}$ into the sum of two terms coming from loop and tadpoles
\begin{align}\label{split}
\Delta W_{S^4} = W_{\rm loop} + W_{\rm tadpole}\,.
\end{align}

\subsection{Contribution of tadpoles}

We start with the contribution of tadpole, which reads
\begin{align} \label{W-W1}
& W_{\rm tadpole}=  -
\frac12{g^4 C_F N} \int_0^{2 \pi} d s_1 \int_0^{2 \pi} d s_2\,
  D_{\rm tadpole}(X_1,X_2) - (m=0) \,.
\end{align}
Replacing $D_{\rm tadpole}$ with its explicit expression \re{D-loop}, we obtain
\begin{align}\label{W-2pt}
W_{\rm tadpole}=  -
\frac12{g^4 C_F N} S_{0} \int_0^{2 \pi} d s_1 \int_0^{2 \pi} d s_2\,[(\dot X_1 \cdot \dot X_2)-|\dot X_1 ||\dot X_2|] I(X_{12}^2)\,,
\end{align}
where $S_0$ is determined by Eq.\ \re{S-tp}. Here we took into account that $S_0$ vanishes for $m=0$ and introduced the notation for the integral
\begin{align}\notag\label{I-2pt}
I(X_{12}^2) &= \int d \sigma_1   D_\Phi (X_1 - Z_1) D_\Phi (X_2 - Z_1)
\\
&={1\over (4\pi)^{2-\epsilon}} \int {dz\over 2\pi i}  \Gamma(-z)\Gamma(-z+\epsilon) \Gamma^2(z+1-\epsilon)(X_{12}^2/4)^z \,,
\end{align}
where the integration contour separates increasing and decreasing poles stemming from $\Gamma(-z+\dots)$ and $\Gamma(z+\dots)$, respectively. The details of the calculation can be found in the Appendix~\ref{app:int}.

We recall that $X_i=X(s_i)$ parameterizes points on the great circle of the sphere (see Eq.~\re{Circle4D}). Substituting \re{I-2pt} into \re{W-2pt}, we  encounter the following contour integrals
\begin{align} \notag
\label{cont-int}
&\int_0^{2\pi} ds_1 ds_2 \left(X_{12}^2/4 \right)^{z} | \dot{X}_1 |  | \dot{X}_2 | = 4 \pi^{3/2} \frac{\Gamma (z+\ft12)}{\Gamma (z + 1)}
\, ,
\\  
&\int_0^{2\pi} ds_1 ds_2 \left( X_{12}^2/4\right)^z (\dot X_1 \cdot \dot X_2) =-4 \pi^{3/2} \frac{  z \Gamma \left(z+\frac{1}{2}\right)}{\Gamma (z+2)}\,.
\end{align}
Notice that these two only differ by a factor of $(-z/(z+1))$ in the right-hand side.
 
Combining the above relations together, we arrive at 
\begin{align} 
W_{\rm tadpole} {}&= 
 {g^4 C_F N\over (4\pi)^{2-\epsilon} } S_{0}  \times 4 \pi^{3/2} 
   \int{dz\over 2\pi i} { \Gamma(-z)\Gamma(\epsilon-z)\Gamma^2(z+1-\epsilon) \Gamma (z+\ft32)  \over \Gamma (z+2)}  \,.
\end{align}     
The integral is finite for $\epsilon = 0$ and it can be expanded at small $\epsilon$ using the {\em MB\,Tools} package \cite{MB}. In this way, we find
\begin{align}\notag\label{G3+S3}
W_{\rm tadpole} {}&= \frac12
 {g^4 C_F N } S_{0}  \lr{\pi \e^{\gamma}}^\epsilon\left(1-\log 2\right) \left[1+2\epsilon  +O (\epsilon ^2 )\right]
 \\
{}& = {g^4 C_F N\over 8\pi^2} \lr{2\pi \e^{\gamma}}^{2\epsilon} m^2\left(1-\log 2\right)\left[ {1\over \epsilon}+ 
   \left(3-H_{i m}-H_{-i m}\right)+O(\epsilon)\right].
\end{align} 
This relation defines the contribution of the sum of the diagrams $G_3$ and $S_3$ in Fig.~\ref{PolOoperatorsFig}, $W_{\rm tadpole}=W_{G_3+S_3}$.
 
\subsection{Contribution of loops} 

In a close analogy with \re{W-W1}, the contribution of graphs containing massive loops is   
\begin{align} \label{W-W2}
& W_{\rm loop}=  -
\frac12{g^4 C_F N} \int_0^{2 \pi} d s_1 \int_0^{2 \pi} d s_2\,
  D_{\rm loop}(X_1,X_2) - (m=0) \,,
\end{align}
where $D_{\rm loop}$ is given by \re{D-loop}.
 
Due to quite an intricate form of the functions involved, a direct closed-form evaluation of the integral in Eq.\ \re{D-loop} is way too complicated. As we show below, these difficulties can be
alleviated by employing a Mellin-Barnes representation of the polarization operators \re{Pi's}
\begin{align}\label{Pi-MB}
\Pi_i(X^2) = \int{dj\over 2\pi i}\, \widetilde \Pi_i(j)\, (X^2/4)^{j} \,.
\end{align} 
In this manner, replacing the scalar propagator in   \re{D-loop} with its expression \re{PhiProp}, we find that the integrals over the sphere in  \re{D-loop} can be expressed in terms of  a `simple' integral 
\begin{align}\label{I-3pt} 
\int  d \sigma_1 d \sigma_2 D_\Phi (X_1 - Z_1) D_\Phi (X_2 - Z_2) (Z_{12}^2/4)^{j}
= \int{dz\over 2\pi i} \widetilde I(j,z) (X_{12}^2/4)^z\,.
\end{align}
Here, the amplitude $\widetilde I(z,j)$ is a meromorphic function of $z$ and $j$
\begin{align}\label{tilde-I}
\widetilde I(j,z) =  
\frac{\Gamma (j-\epsilon +2)\Gamma (-z) \Gamma (j-z+2) \Gamma (z-j) \Gamma ^2(z-\epsilon +1)}{\Gamma (-j) \Gamma^2 (j-\epsilon +3)\Gamma
   (z-\epsilon +2)}\,.
\end{align}
Its derivation can be found in Appendix~\ref{app:int},  (see Eq.~\re{I-J}).

Combining together the above relations, we get from \re{W-W2} the representation of $W_{\rm loop}$ as a double Mellin-Barnes integral
\begin{align}\notag\label{W-multi}
W_{\rm loop}& =  2{g^4 C_F N}  \int{dj\over 2\pi i}\,  \int{dz\over 2\pi i} \widetilde I(j,z) 
\\
& \times \int_0^{2 \pi} d s_1 \int_0^{2 \pi} d s_2\,   
\left[(\dot X_1 \cdot \dot X_2) \widetilde \Pi_1(j) + |\dot X_1 ||\dot X_2|  \widetilde \Pi_2(j)  
\right](X_{12}^2/4)^z - (m=0)\,,
\end{align}
where the amplitudes $\widetilde \Pi_1(j)$ and $\widetilde \Pi_2(j)$ can be obtained from \re{Pi-MB} and \re{Pi's} by replacing the functions $A$, $B$ and $S_\pm$ by their Mellin-Barnes 
images (see Eqs.~\re{S-MB} and \re{A-MB}).

The contour integrals in the second line of \re{W-multi} can be evaluated using \re{cont-int}. This leads to
\begin{align}\label{W-MB}
W_{\rm loop}& = -8 \pi^{3/2}
 {g^4 C_F N}   \int{dz\over 2\pi i}\,   \frac{\Gamma (z+\ft12)}{\Gamma (z + 1)}
\left[ { z\over z+1}\mathcal M_1(z) - \mathcal M_2(z)\right]  - (m=0)\,,
\end{align}
where the notations were introduced for 
\begin{align}\notag\label{M's}
\mathcal M_1(z) &= \int{dj\over 2\pi i}\, \widetilde I(j,z)  \widetilde \Pi_1(j) = \mathcal M_A-\mathcal M_B\,,
\\
\mathcal M_2(z) &= \int{dj\over 2\pi i}\, \widetilde I(j,z)  \widetilde \Pi_2(j) = \mathcal M_A+\mathcal M_B + \mathcal M_S\,.
\end{align}
Here the functions $\mathcal M_A$, $\mathcal M_B$ and $\mathcal M_S$ describe contributions of
various terms in the right-hand side of Eq.\ \re{Pi's}. Their explicit expressions in terms of the Mellin-Barnes amplitudes of
the functions $A$, $B$ and $S_\pm$ are
\begin{align}\notag
& \mathcal M_A=4 \int {dz_1 dz_2\over (2\pi i)^2} \widetilde A(z_1)\widetilde A(z_2)\left[\widetilde I(z_1+z_2,z) -\widetilde I(z_1+z_2+1,z)   \right] \,,
\\\notag
& \mathcal M_B=4 \int {dz_1 dz_2\over (2\pi i)^2} \widetilde B(z_1)\widetilde B(z_2) \widetilde I(z_1+z_2+1,z)  \,,
\\
& \mathcal M_S=m^2 \int {dz_1 dz_2\over (2\pi i)^2} \left[ \widetilde S_+(z_1)\widetilde S_+(z_2) +\widetilde S_-(z_1)\widetilde S_-(z_2) \right]\widetilde I(z_1+z_2,z)\,,
\end{align}
where $\widetilde I$ is given by \re{I-3pt} while $\widetilde S_\pm$, $\widetilde A$ and $\widetilde B$ are determined in Eqs.\ \re{tildeS} and \re{A-MB1}.

It is convenient to split the relation \re{W-MB} into the sum of contributions of individual graphs shown in Fig.~\ref{PolOoperatorsFig}
\begin{align}\label{loop-sum}
W_{\rm loop} = W_{ G_1+G_2} +  W_{ S_1} + W_{ S_2} -(m=0)  \,.
\end{align}
We recall that the sum of graphs $(G_1)$ and $(G_2)$ yields $\Pi_1$ in the first equation of \re{Pi's} and, therefore, it produces the contribution to \re{W-MB} proportional to $\mathcal M_1$. 
The diagram $(S_1)$ generates the first two terms in the expression for $\Pi_2$ in \re{Pi's} and its contribution to \re{W-MB} is described by the sum of two terms $\mathcal M_A+\mathcal M_B$ 
in the expression for $\mathcal M_2$ in \re{M's}. The remaining $\mathcal M_S$ term in \re{M's} comes from the graph $(S_2)$. 

The calculation of the contribution to \re{W-MB} from $\mathcal M_B$ term in the expression for $\mathcal M_2$ is presented in Appendix~\ref{app:S1}. 
The contribution of the remaining terms can be found in a similar manner. The resulting expressions for the individual graphs in \re{loop-sum} take the form of (complicated) triple 
Mellin-Barnes integrals. In the next subsection, we present their expansion in the limit of small and large masses. For finite mass, we used the {\em MB\,Tools} package \cite{MB} 
to compute them numerically.

\subsection{Two loops versus localization}

In this subsection, we summarize the obtained results and present the two-loop expressions for the circular Wilson loop 
\begin{align}\label{sum-W}
\Delta W_{S^4} & = \sum_\alpha \lr{W_\alpha(m)-W_\alpha(0)} \,,
\end{align}
where $\alpha=\{G_1+G_2,S_1,S_2,G_3+S_3\}$ enumerates the graphs  shown in Fig.~\ref{PolOoperatorsFig}.

In the small mass limit, the individual contributions of the diagrams admit the following general form
\begin{align}\label{W-small}
W_\alpha(m)= - {g^4 C_F N\over 8\pi^2}\left[w_\alpha^{(0)} + m^2 w_\alpha^{(1)} + O(m^4) \right],
\end{align}
 Explicit expressions for the leading coefficients $w_\alpha^{(0)}$  read
\begin{align}\notag
{}& w^{(0)}_{G_1+G_2} =(\pi \e^{\gamma})^{2\epsilon} \left[{1\over 4\epsilon}+1 +O(\epsilon) \right]
\,,
\\\notag
{}& w^{(0)}_{S_1} =-\frac12+O(\epsilon) 
\,,
\\[2mm]
{}& w^{(0)}_{S_2}  = w^{(0)}_{G_3+S_3}= 0\,.
\end{align}
The subleading coefficients are given by 
\begin{align}\notag
  {}&w^{(1)}_{G_1+G_2} =(2\pi \e^{\gamma})^{2\epsilon} \left[\frac{1-2\log 2}{2 \epsilon }+  \left(-\frac32 \zeta (3)+\frac32+2\log 2\right)+O\left(\epsilon
   \right)\right],
\\\notag
  {}&w^{(1)}_{S_1} =(2\pi \e^{\gamma})^{2\epsilon} \left[ \frac{3}{2 \epsilon }+\left(\frac{11}{2}-11 \log 2\right)+O\left(\epsilon\right) \right] ,
\\\notag
  {}&w^{(1)}_{S_2} =(2\pi \e^{\gamma})^{2\epsilon} \left[-\frac{1}{\epsilon }+(6\log 2-4)+O\left(\epsilon\right) \right],
\\
  {}&w^{(1)}_{G_3+S_3} =(2\pi \e^{\gamma})^{2\epsilon} \left[ \frac{\log 2-1}{\epsilon }+(3 \log 2-3)+O\left(\epsilon\right)\right],
\end{align}
where $w^{(1)}_{G_3+S_3}$ was obtained by expanding the tadpole contribution \re{G3+S3} at small $m$.
Substituting \re{W-small} into \re{sum-W}, we find in the limit $\epsilon\to 0$
\begin{align} 
\Delta W_{S^4}  
& =- {g^4 C_F N\over 4\pi^2}  m^2  \sum_\alpha w_\alpha^{(1)}  
=  {g^4 C_F N\over 16\pi^2} \left[ 3\zeta(3)m^2 + O(m^4)\right]\,.
\end{align}
In agreement with our expectations, UV poles cancel in the sum of all graphs. Moreover, finite rational terms and terms proportional to $\log 2$ also cancel 
against each other as well. The resulting expression for $\Delta W_{S^4}$ coincides with the localization prediction \re{diff} and \re{loc-small}.
 
In the opposite large-mass limit, the contribution of the graphs in Fig.~\ref{PolOoperatorsFig}  looks like
\begin{align}\label{W-large}
W_\alpha(m)= - {g^4 C_F N\over 8\pi^2}\lr{\pi\over m}^{2\epsilon} w_\alpha^{(\infty)}\,,
\end{align}
with the explicit expressions for the coefficients being
\begin{align}\notag
 {}& w^{(\infty)}_{G_1+G_2} = {1\over\epsilon}\left(m^2 {{\left(1-2\log 2\right)}\over 2} +\frac14 \right) +m^2
   \left(\frac{3}{2}-2 \log ^2 2-\log 2\right)
\\ &    \notag
 \qqqquad 
   + \frac{\pi m}2+\frac{5+2\log 2}{12} -{\pi\over 32m} +{2+\log 2\over 60m^2} - {11\pi\over 2048m^3}+ \frac{43+40 \log (2)}{5040 m^4}+O\left(\epsilon\right)\,,
\\\notag
 {}&w^{(\infty)}_{S_1} =\frac{3 m^2}{2 \epsilon }+\frac{7 m^2}{2}-\frac{3 \pi 
   m}{4}-\frac{3}{4} + {9\pi\over 128m}-{1\over 40m^2} +\frac{63 \pi }{4096 m^3}-\frac{1}{84 m^4}+O\left(\epsilon \right)\,,
\\\notag
 {}&w^{(\infty)}_{S_2} =   -\frac{m^2}{\epsilon }-2 m^2+\frac{\pi  m}{4}+\frac{1}{6}-{5\pi\over 128m} +{1\over 60m^2}-\frac{41 \pi }{4096 m^3}+\frac{1}{126 m^4}+O\left(\epsilon
   \right)\,,
\\\notag
 {}&w^{(\infty)}_{G_3+S_3} = m^2\frac{(\log 2-1)}{\epsilon }+m^2 \left(-3+2 \log ^2 2+\log
   2\right)
\\
{}& \hspace*{16mm}   
   +\frac{1-\log 2}{6}+\frac{1-\log 2}{60m^2} +\frac{1-\log (2)}{126 m^4}+O\left(\epsilon\right)\,.
\end{align}
The sum of the coefficients takes a remarkably simple form
\begin{align}
\sum_\alpha w^{(\infty)}_{\alpha}={1\over 4\epsilon} + {1\over 24m^2}+\frac{1}{80 m^4} + O(\epsilon)\,.
\end{align}
Substituting \re{W-large} into \re{sum-W}  and using this relation, we obtain the circular Wilson loop at large $m$ as
\begin{align} \notag
\Delta W_{S^4} 
&
= - {g^4 C_F N\over 8\pi^2}  \sum_\alpha \left[{\lr{\pi\over m}^{2\epsilon}w_\alpha^{(\infty)}  -w_\alpha^{(0)}}\right]
\\
& = {g^4 C_F N\over 16\pi^2}\left(\log m + \gamma+1 -{1\over 12m^2}-\frac{1}{40 m^4} +O\left({1\over m^6}\right)\right).
\end{align}
As in the previous case, we observe perfect agreement with the localization prediction, Eqs.~\re{diff} and \re{loc-large}.
 
For finite values of the mass, we computed \re{sum-W} numerically for various values of $m$ and reproduced the expected result \re{diff} and \re{LocalizationFiniteM} to high precision.
 
\section{Conclusions}
\label{SectionConclusions}

In this paper, we developed a framework that allowed us to compute the circular Wilson loop in the $\mathcal N=2^\ast$ super Yang-Mills on the four-sphere at two loop
order. We verified that it perfectly agrees with the prediction of the supersymmetric localization but differs from analogous perturbative calculations in the flat space. In the 
latter case, the difference arises due to the presence of a mass scale in the theory. The reason being that the mass deformation explicitly breaks the conformal symmetry 
of the model and, as a consequence, a coordinate transformation from the sphere to the flat space becomes anomalous. This question deserves further investigation.

The main findings of our work are rather technical but they have a potential to be of value in other circumstances as well. We demonstrated that the calculation on the sphere 
can be simplified by employing the embedding coordinates instead of local coordinates (like spherical angles or stereographic coordinates). We derived, in particular, a new, 
very concise representation for a massive fermion propagator which proves to be very convenient in evaluating Feynman integrals on the sphere. 

Having worked in the difference theory, we encountered two-loop graphs of propagator type only, that is, one-loop graphs with massless propagators modified by corrections 
from massive fields circulating in virtual loops. As a first step in computing these graphs, we proposed to use the Mellin-Barnes representation for the massive propagators.
The main advantage of this representation is that it allows us to disentangle the dependence on the mass and the chordal distance on the sphere and, as a consequence, to 
reduce our calculations merely to the evaluation of integrals on the sphere containing products of powers of chordal distances. With all integrations done, we arrived at a compact 
representation for the two-loop circular Wilson loop as a sum of three-fold Mellin-Barnes integrals involving ratios of Euler $\Gamma-$functions. These integrals are performed
in the complex plane along the contours separating increasing and decreasing sequences of poles generated by the latter.
 
Using the Mellin-Barnes representation as a starting point, we can apply conventional techniques, reviewed in Ref.\ \cite{Smirnov:2012gma}, and compute emerging integrals 
by residues. We did not succeed in getting their closed-form expressions for finite value of the mass since a brute force use of Cauchy theorem is unwieldy and results in nested 
infinite series representation that we failed to resum. This is the reason why we limited ourselves only to their small and large mass expansions. For finite value of the mass, however,
we computed contributing Mellin-Barnes integrals numerically to high precision. In all cases, we found that the two-loop result for the circular Wilson loop is in an exact agreement with 
the localization. Our consideration can thus be regarded as a first test of the latter in massive non-conformal settings. 

The eventual two-loop result for the circular Wilson loop arises after massive cancellations between contributing two-loop graphs. This observation, as well as the simplicity of the localization 
formulae, hints that the Mellin-Barnes integrals mentioned above can be computed exactly for finite value of the mass as well. It would be interesting to observe this explicitly.

The formalism developed in this work can be used in computation of other observables in massive super Yang-Mills theories on the four-sphere, e.g., correlation functions of local 
operators or Wilson loops with insertions of these. Also, since de Sitter space can be obtained in the embedding coordinates by a Wick rotation, all ingredients of our perturbative 
construction can be used in cosmological applications as well.

\section*{ Acknowledgments} 

We are grateful to Marco Billo, Francesco Fucito, Alberto Lerda and Francisco Morales for collaboration at the early stage of this project.
We would also like to thank Luca Griguolo, Michelangelo Preti and Domenico Seminara for useful discussions. The research of A.B.\ was supported 
by the U.S. National Science Foundation under the grant PHY-1713125 and of G.K.\ by the French National Agency for Research grant ANR-17-CE31-0001-01.

\appendix

\section{Local coordinates on the sphere}\label{app:loc}

The spherical coordinates on the sphere $X^2=R^2$ are defined as
 \begin{align}
\label{SphericalCoordinatesSD}
\begin{array}{ll}
X_{0} &= R \cos \theta_{1}
\, , \\ [2mm]
X_{1} &= R \sin \theta_{1} \cos \theta_{2}
\, , \\ [2mm]
X_{2} &= R \sin \theta_{1} \sin \theta_{2} \sin \theta_{3}
\, , \\ [2mm]
\ \vdots &
\\ [2mm]
X_{D} &= R \sin \theta_{1} \sin \theta_{2} \sin \theta_{3} \dots \sin \theta_{D-1} \sin \theta_D
\, .
\end{array}
\end{align}
The volume element then reads
\begin{align}
d \sigma \equiv \sqrt{\mbox{\sl g}} \, d^D x = R^D d \Omega^{(D+1)}
\, ,
\end{align}
where $d \Omega^{(D+1)}$ is the element of the solid angle in the above spherical coordinates. The total volume of $S^{D}$ is
\begin{align}
\label{VolSD}
{\rm vol} \, S^D \equiv \int d\sigma = \frac{2 \pi^{\ft{D+1}{2}} }{\Gamma \left( \ft{D+1}{2} \right)}R^D
\, .
\end{align}

\section{Scalar propagator}
\label{AppendixScalarPropagator}

In this Appendix, we derive the massive scalar propagator on the sphere \re{D-AB} and \re{PropAandBscalars}.

As was mentioned in Section \ref{SectionEmbedding}, the calculation can be simplified by Wick rotating to the de Sitter space and using the coordinates \re{EmbedToConfTime}. 
The de Sitter space is conformally flat and its metric can be cast into the form
\begin{align}
g_{\mu\nu} = (z^0)^{- 2} \eta_{\mu\nu}\, ,\qqqquad g = \det g_{\mu\nu} = - (z^0)^{- 8}\,,
\end{align}
with $\eta_{\mu\nu}={\rm diag}\ (+,-,\dots,-)$ and $z^0$ being the conformal time \re{EmbedToConfTimeInverse}. The Laplace-Beltrami operator admits the form
\begin{align}
\label{LBoperator}
\nabla^2 \equiv \nabla^\mu \nabla_\mu
= 
\frac{1}{\sqrt{g}} \partial_\mu \sqrt{g} g^{\mu\nu} \partial_\nu
=
(z^0)^2 (\partial_0^2 - \partial_i^2) - 2 z^0 \partial_0
\, .
\end{align}
The propagator of a conformally coupled real scalar of mass $\mu$ obeys the equation
\begin{align}
\left( - \nabla^2_1 + \ft14 D (D-2) + \mu^2 \right) G_\mu (z_1, z_2) = \sqrt{- g_1} \delta^{(D)} (z_1 - z_2)
\, .
\end{align}
The function $G_\mu (z_1, z_2)$ depends on the coordinates only through the chordal distance,
\begin{align}
\label{ChordalDistanceDS}
s(z_1,z_2) \equiv \frac{-(z_{12}^0)^2 + ({z}_{12}^i)^2}{z^0_1 z^0_2}
\, .
\end{align}
Then, for $z_1 - z_2 \neq 0$ the above equation reduces to 
\begin{align}
\label{ScalarGreenEq}
\left[ s (s - 4) \partial_s^2 + D (s - 2) \partial_s + \ft14 D (D - 2) + \mu^2 \right] G_\mu (s) = 0
\, .
\end{align}
The solution to this equation yields the well-known expression for the massive scalar propagator~\cite{Chernikov:1968zm,Dowker:1975tf,Candelas:1975du}
\begin{align}
\label{CRmassiveProp}
G_\mu (s) 
= 
{\Gamma( \ft12(D-1+M)) \Gamma(\ft12(D-1-M)) \over (4\pi)^{D/2} \Gamma(\ft{D}{2}) }
\left. {}_2 F_1 \left( {\ft12(D-1+M), \ft12(D-1-M) \atop \ft12 D} \right|1 - \frac{s}{4} \right),
\end{align}
where a notation was introduced for $M = \sqrt{1-4 \mu^2}$. In the vanishing mass limit, i.e., for $\mu\to 0$, we obtain
\begin{align}
G_0(s) = \frac{\Gamma (\ft{D}2 -1)}{4 \pi^{D/2}} \frac{1}{s^{D/2-1}}
\, ,
\end{align}
After the analytic continuation, for $s = X_{12}^2$, this relation yields the massless scalar propagator on the sphere  \re{PhiProp}.

The the $\mathcal N=2^\ast$ theory contains two pairs of scalars with masses $\mu_\pm$, see Eq.\ \re{AandBmasses}. Substitution of $\mu=\mu_\pm$ into \re{CRmassiveProp} 
yields the two propagators (with $M = 1\pm 2im$)
\begin{align}
G_{\mu_\pm} (s)=
\frac{\Gamma (\ft{D}{2} - 1 \pm i m) \Gamma (\ft{D}{2} \mp i m)}{(4 \pi)^{D/2} \Gamma (\ft{D}{2})}\,
{_2 F_1}\left.\left({\ft12 D - 1 \pm i m\, , \ft12 D \mp i m\atop\ft12 D}
\right| 1 - \frac{s}{4}\right). 
\end{align}
Upon the analytic continuation to the sphere, these provide the propagators for $A$ and $B$ scalars introduced in \re{D-AB} with $S_\pm(X_{12}^2) = G_{\mu_\pm} (s)$ for $s=X_{12}^2$.

\section{Fermion propagator}
\label{AppendixFermionPropagator}

In this appendix, we derive the massive fermion propagator \re{ferm-prop}.~\footnote{We provide a comprehensive analysis since we failed to find it in the existing literature.} 
We do it a step-wise manner, first, by finding the propagator in the de Sitter coordinates \re{EmbedToConfTimeInverse}, and then, lifting it to the embedding space. It is the latter 
form that we use in the main body. 

The propagator of the fermion with the mass $m$ obeys the following equation 
\begin{align}\label{Dirac}
(i {\slashed\nabla}_1 - m) S_m (z_1, z_2) = \frac{1}{\sqrt{- g_1}} \delta^{(D)} (z_1 - z_2)
\, ,
\end{align}
where the Dirac operator can be written in terms of vielbeins as
\begin{align}\label{nabla}
{\slashed\nabla} = \gamma^a e^\mu_a
\left(
\frac{\partial}{\partial z^\mu} + \frac12 \Sigma_{ab} \,\omega^{ab}_\mu
\right),
\end{align}
with the spin matrix and the spin connection given by
\begin{align}
\Sigma_{ab} = \ft14 [\gamma_a, \gamma_b]
\, ,\qqqquad
\omega^{ab}_\mu
=
e^{\nu,a}
\left(
\frac{\partial}{\partial z^\mu} e^b_\nu - \Gamma^\rho_{\nu\mu} e^b_\rho
\right)
\, .
\end{align}
They involve the flat-space Dirac matrices obeying the Clifford algebra $\{\gamma_a, \gamma_b\} = 2 \eta_{ab}$ and the affine connection
\begin{align}
\Gamma^\lambda_{\rho\mu}
=
\frac12 g^{\lambda\nu}
\left(
\frac{\partial}{\partial z^\rho} g_{\mu\nu}
+
\frac{\partial}{\partial z^\mu} g_{\rho\nu}
-
\frac{\partial}{\partial z^\nu} g_{\mu\rho}
\right).
\end{align}
In the conformally-flat de Sitter metric, its only nonvanishing components are
\begin{align}
\Gamma^0_{ij} = \frac{\delta_{ij}}{z^0}
\, , \qqquad
\Gamma^i_{0j} = - \frac{\delta_{ij}}{z^0}
\, , \qqquad
\Gamma^i_{j0} = - \frac{\delta_{ij}}{z^0}
\, , \qqquad
\Gamma^0_{00} = - \frac{1}{z^0}
\, .
\end{align}
The de Sitter metric in terms of vielbeins looks as
\begin{align}\label{g-co}
g_{\mu\nu} = e_\mu^a e_{\nu, a}\,, 
\qqquad
e_\mu^a = (z^0)^{-1} \delta_\mu^a
\, , \qqquad 
e_{\nu, a} = (z^0)^{-1} \eta_{\nu a}
\, ,
\end{align}
where the Greek labels denote the curved space-time indices and the Latin ones stand for the flat space-time indices of a local 
inertial frame. The inverse metric is then
\begin{align}
\label{InvVielebein}
g^{\mu\nu} = e^\mu_a e^{\nu, a}
\, ,
\qqquad
e^\mu_a = z^0 \delta^\mu_a
\, , \qqquad 
e^{\nu, a} = z^0 \eta^{\nu a}
\, .
\end{align}
The spin connection possesses the following nonzero elements
\begin{align}
\omega^{0i}_j = - \omega^{i0}_j = \frac{\delta_{ij}}{z^0}
\, ,
\end{align}
such that the Dirac operator explicitly reads
\begin{align}
\slashed\nabla
=
\gamma^0 
\left(
z^0 \frac{\partial}{\partial z^0} - \frac{D - 1}{2}
\right)
-
z^0 \gamma^i \frac{\partial}{\partial z^i} 
=
(z^0)^{(D + 1)/2} {\not\!\partial} (z^0)^{- (D -1)/2}
\, .
\end{align}

Then, the solution to \re{Dirac}  has the form
\begin{align}\label{FermionPropConfTime}
S_m (z_1, z_2) = \frac{1}{z_1^0} (i {\slashed\nabla}_1 + m) \lr{z_1^0 z_2^0}^{1/2} 
\left[ 
\frac{1 + \gamma^0}{2} S_+  (s) + \frac{1 - \gamma^0}{2}  S_- (s)
\right]
\, ,
\end{align}
where $S_\pm(s)$ are the functions of the chordal distance \re{ChordalDistanceDS} defined in Eq.\ \re{PropAandBscalars}.
To verify this relation, we use the identity
\begin{align}
- \sqrt{z_1^0} (i {\slashed\nabla}_1 - m) \frac{1}{z_1^0} (i {\slashed\nabla}_1 + m) \sqrt{z_1^0}
=
s (s - 4) \partial_s^2 + D (s - 2) \partial_s + \ft14 D (D - 2) + m^2 + i m \gamma^0
\, .
\end{align}
Notice that the first three terms in its right-hand side coincide with those inside the brackets in Eq.\ \re{ScalarGreenEq}. Substituting \re{FermionPropConfTime} into \re{Dirac}, 
we find that $S_\pm(s)$ obey Eq.\ \re{ScalarGreenEq} with $\mu_\pm^2=m^2\pm im$. Since $S_\pm=G_{\mu_\pm}$ these equations are automatically verified.

Expanding the expression in the right-hand side of Eq.\ \re{FermionPropConfTime} and taking into account the relation \re{difS}, we find after some algebra 
\begin{align}\notag\label{Sm-int}
i S_m (z_1, z_2)
={}&
-
\frac{{\not\!{\widetilde z}}_{12} \gamma^0 }{\sqrt{z_1^0 z_2^0}} 
\frac{(\ft12 D - 1 + i m ) S_+ (s)-(\ft12 D - 1 - i m ) S_- (s)}{s - 4} 
\\
&-\frac{{\not\!z}_{12}}{\sqrt{z_1^0 z_2^0}} \frac{
(\ft12 D - 1 + i m ) S_+ (s)+(\ft12 D - 1 - i m ) S_- (s)}{s} 
\, ,
\end{align}
where a notation was introduced for $\not\!{\widetilde{z}}_{12} = (z_1^0+z_2^0) \gamma^0 -(\bit{z_1-z_2})\cdot\bit{\gamma}$ and
$\not\!{{z}}_{12} =\not\!{{z}}_{1}-\not\!{{z}}_{2}$ with $\not\!{{z}}_{i} = z_i^0\gamma^0-\bit{z}_i \cdot\bit{\gamma}$ and a convention used 
for $\bit{\gamma} \cdot \bit{z}=\sum_{i=1}^{D-1}\gamma^i z^i$.

The representation of the fermion propagator in the form \re{FermionPropConfTime} has been known for quite some time from Ref.\ \cite{Candelas:1975du}. 
Our next goal is to obtain an analogous, covariant representation of this propagator in the embedding coordinates.

Let us transform the fermion propagator \re{FermionPropConfTime} from the local $z$-coordinates to the embedding $X$-coordinates \re{EmbedToConfTimeInverse}. To this end, we
introduce the matrix
\begin{align}
U(z) = \frac{1}{2 \sqrt{z^0}}
\left[
(1 + z^0) + \gamma^0 \gamma^D (1 - z^0)-\left( \gamma^0 + \gamma^D \right) \bit{\gamma} \cdot \bit{z}\right]\, ,
\end{align}
where $\gamma^M$ (with $M=0,\dots,D)$ are the Dirac matrices in $(D+1)$-dimensional flat space-time with the signature $\eta^{MN} = {\rm diag} (+,-,\dots,-)$. The inverse matrix looks as 
 \begin{align}
U^{-1}(z) = \frac{1}{2 \sqrt{z^0}}\left[(1 + z^0)- \gamma^0 \gamma^D (1 - z^0)+\left( \gamma^0 + \gamma^D \right) \bit{\gamma} \cdot \bit{z}\right]\, .
\end{align}
A simple calculation shows that
\begin{align} 
& U \gamma^0 U^{-1} 
= z^0 \gamma^M {\partial X_M\over\partial z^0}
\,,\qquad
 U \gamma^i U^{-1}
= z^0 \gamma^M {\partial X_M\over\partial z^i}
\,,\qquad
U \gamma^D U^{-1} 
= \gamma^M X_M\,,
\end{align}
where the embedding coordinates $X_M$ are given by \re{EmbedToConfTime} for $R=1$. In the covariant form, these relations were presented in the main text in Eq.\ \re{GammaRotations}.

It is straightforward to verify that
\begin{align}\notag
& U_1 \gamma^a z_{12,a} U_2^{-1}=\lr{z_1^0 z_2^0}^{1/2} \ \gamma^M X_{12,M}\,,
\\
& U_1{\gamma^a \gamma^0 \widetilde{z}_{12,a}}  U_2^{-1} =- \lr{z_1^0 z_2^0}^{1/2} \gamma^M  \gamma^N \widetilde{X}_{12,M}  X_{2,N}
\,,
\end{align}
where $U_i \equiv U(z_i)$ and $ \widetilde{X}_{12} \equiv {X}_{1}+ {X}_{2}$. Combining these relations with \re{Sm-int}, we finally obtain
\begin{align}
i S _m(z_1, z_2)
=
U_1^{-1}
\bigg[
&
\frac{(\ft12 D - 1 + i m ) S_+ (s)-(\ft12 D - 1 - i m ) S_- (s)}{s - 4} {\not\!\!\widetilde{X}}_{12}  {\not\!\!X}_2
\nonumber\\
&
-\frac{(\ft12 D - 1 + i m ) S_+ (s)+(\ft12 D - 1 - i m ) S_- (s)}{s} {\not\!\!X}_{12}\bigg]U_2\, .
\end{align}
After the analytical continuation from the de Sitter space to the sphere, this relation yields the propagator of the massive fermion quoted in Eq.\ \re{ferm-prop}.

\section{Mellin-Barnes representation of massive propagators}\label{app:MB}

In this appendix, we discuss the properties of the functions $S_\pm$, $A$ and $B$ defining the propagators of the massive scalars \re{D-AB} and fermions \re{ferm-prop}. 

These functions are not independent and are related to each other by the linear relations \re{Bfermion}. The inverse transformation looks like 
\begin{align}\label{SfromAB}
S_+(s) = {(s-4)A(s) + s B(s) \over D-2+2im}\,,\qqquad 
S_-(s) = {-(s-4)A(s) + s B(s) \over D-2-2im}\,,
\end{align}
where $s=X^2$ and $D=4-2\epsilon$.

We start with the functions $S_\pm(X^2)$ given by Eq.\ \re{PropAandBscalars}. Using the properties of the hypergeometric functions (see Eq.~\re{2MBf21}), we can immediately obtain 
its Mellin-Barnes representation
\begin{align}\label{S-MB}
S_\pm (X^2) {}& = \int {dz \over 2\pi i}  (X^2/4)^z \, \widetilde S_\pm(z)\,,
\end{align}
where its image is given by 
\begin{align}\label{tildeS}
\widetilde S_\pm(z) = \frac{  \Gamma (-z) \Gamma (-z-1+\epsilon )\Gamma (z+1\pm i m-\epsilon ) \Gamma
   (z+2\mp i m-\epsilon ) }{(4\pi)^{2-\epsilon} \Gamma (\pm i m) \Gamma (1\mp i m) }\,.
\end{align}
The integration contour in \re{S-MB} separates increasing and decreasing poles generated by $\Gamma-$functions of the form $\Gamma(-z+\dots)$ and $\Gamma(z+\dots)$, respectively. 
The main advantage of this form is that the integrand has a power-like dependence on the chordal distance $X^2$ and is better suited for performing integration on the sphere.
 
Substituting \re{PropAandBscalars} into  \re{Bfermion}, we can re-express the functions $A(X^2)$ and $B(X^2)$ in terms of the hypergeometric functions
\begin{align}\notag\label{AB-hyper}
A(X^2){}&=- {i m } {\Gamma(D/2+im)\Gamma(D/2-im)\over  2^{D+1} \pi^{D/2}\Gamma(D/2+1)}
{}_2F_1 \lr{{D/2+im,D/2+im\atop D/2+1} \Big| 1-X^2/4}\,,
\\[2mm]
B(X^2){}& = {\Gamma(D/2+im)\Gamma(D/2-im)\over 2^{D+1} \pi^{D/2} \Gamma(D/2)}
{}_2F_1 \lr{{D/2+im,D/2+im\atop D/2} \Big| 1-X^2/4}\,.
\end{align}
It is obvious from the first relation that $A(X^2)$ vanishes as $m\to 0$.

As in the previous case, we can work out the Mellin-Barnes representation of the functions \re{AB-hyper}
\begin{align}\label{A-MB}
A(X^2) {}&=   \int {dz \over 2\pi i}  (X^2/4)^z \, \widetilde A(z)
\,,\qqquad
B(X^2)  =\int {dz \over 2\pi i}  (X^2/4)^z \, \widetilde B(z)\,,
\end{align}
where their amplitudes now read
\begin{align}\notag\label{A-MB1}
\widetilde A(z) {}&= - {i m }  {\Gamma(-z)\Gamma(-z-1+\epsilon)\Gamma(2+im+z-\epsilon)\Gamma(2-im+z-\epsilon)\over 2(4\pi)^{2-\epsilon}\Gamma(1+im)\Gamma(1-im)}  \,,
\\
\widetilde B(z) {}& = {\Gamma(-z)\Gamma(-z-2+\epsilon)\Gamma(2+im+z-\epsilon)\Gamma(2-im+z-\epsilon)\over  2(4\pi)^{2-\epsilon}\Gamma(im)\Gamma(-im)}  \,.
\end{align}
Again, the integration contour in \re{A-MB} separates increasing and decreasing poles generated by $\Gamma(-z+\dots)$ and $\Gamma(z+\dots)$, respectively. Note that the two amplitudes 
differ by a simple factor,  $\widetilde A(z)/\widetilde B(z)=(-z-2+\epsilon)/(im)$.   The relation \re{SfromAB} translates to the analogous relation between their Mellin-Barnes images
\begin{align}
\widetilde S_\pm(z) = 2 { \widetilde B(z-1)\pm (\widetilde A(z-1) - \widetilde A(z)) \over 1\pm im -\epsilon}\,.
\end{align}
We can apply \re{A-MB} to show that the functions satisfy the following differential equations      
\begin{align}
A'(s) = {D A(s) + 2im B(s) \over 2(4-s)}\,,\qqquad
B'(s) = - {D B(s) + 2im A(s) \over 2s}\,.
\end{align}
Combining them with \re{SfromAB}, we verify that
\begin{align}\notag
{}& S'_\pm (s) = \frac{2 (\ft12 D - 1 \mp i m) }{s (s - 4)} S_\mp (s) 
+ \frac{(2-s ) (\ft12 D - 1 \pm i m) }{s (s - 4)} S_\pm (s)\,,
\\
\label{difS}
{}& (S_+'(s))^2 + (S_-'(s))^2 = \frac12(A^2(s) + B^2(s))\,.
\end{align}
The contribution of tadpole graphs involves the function $A(X^2)$ evaluated at coincident points $X^2=0$. It can be found from the first relation in \re{A-MB} by moving the integration 
contour to the right and picking up the residue at $z=0$
\begin{align}\label{app-A0}
A(0) = -\frac{i m \Gamma (\epsilon-1) \Gamma (2-\epsilon-i m) \Gamma (2-\epsilon+i m)}{2(4\pi)^{2-\epsilon}\Gamma (1-i m) \Gamma (1+i m)}\,.
\end{align}

\section{Integrals on the sphere}\label{app:int}

In this appendix, we evaluate integrals on the sphere which appear in the calculation of the circular Wilson loop. 

We begin with the simplest integral
\begin{align}\label{J1}
J(\nu)=\int {d \sigma \over |X-Z|^{2\nu}} 
\, ,
\end{align}
where the integration measure is defined in \re{measure} and $ |X-Z|^{2\nu}=[(X-Z)^2]^\nu$. In virtue of the rotational invariance, $J(\nu)$ does not depend on the choice of $X$ on the 
sphere. Choosing  $X = (1,0,\dots, 0)$ and passing to the spherical coordinates \re{SphericalCoordinatesSD}, we get $(X-Z)^2 = 2 (1 - \cos\theta_1)$. Replacing the integration measure with
\begin{align}
\label{OmegaDplus1}
d \sigma=d \Omega^{(D+1)} = d \Omega^{(D)} d \theta_1 \sin^{D-1}\theta_1 \,,
\end{align}
we get from \re{J1} 
\begin{align}
\label{J1}
J(\nu)=  2^{D-2\nu-1} {\Gamma(D/2)\Gamma(D/2-\nu)\over \Gamma(D-\nu)}  \Omega^{(D)}
=\frac{(4\pi) ^{D/2} \Gamma \left( {D}/{2}-\nu \right)}{2^{2 \nu } \Gamma (D-\nu )}
\, .
\end{align}
Setting $\nu = 0$, we verify that $J(0)$ reduces to Eq.\ \re{VolSD}.

Next, let us consider a more complicated integral
\begin{align}\label{J2-def}
J (\nu_1, \nu_2)= {}&  \int {d\sigma \over |X_1-Z|^{2\nu_1}|X_2-Z|^{2\nu_2}} \,.
\end{align}
Taking into account that $|X_i-Z|^{2\nu_i} = [2(1-(X_i Z))]^{\nu_i}$ and using the conventional Feynman parametrization, we get
\begin{align}
{1 \over |X_1-Z|^{2\nu_1}|X_2-Z|^{2\nu_2}} = {\Gamma(\nu_1+\nu_2)\over \Gamma(\nu_1)\Gamma(\nu_2)}
\int_0^1 {dy\, y^{\nu_1-1}(1-y)^{\nu_2-1}\over [2(1-|V(y)| \cos\theta)]^{\nu_1+\nu_2}}\,,
\end{align}
where $\theta$ is the angle between $Z$ and the vector $V(y)= y X_1 + (1-y)X_2$ with length $|V(y)|=[1-y(1-y)X_{12}^2]^{1/2}$.
Applying \re{OmegaDplus1} and changing the integration variable to $x=\cos\theta$, we obtain
\begin{align}
J (\nu_1, \nu_2)= \Omega^{(D)} {\Gamma(\nu_1+\nu_2)\over \Gamma(\nu_1)\Gamma(\nu_2)} \int_{-1}^1 dy\, y^{\nu_1-1}(1-y)^{\nu_2-1} 
\int_0^1 {dx (1-x^2)^{D/2-1} \over [2(1- x |V(y)|)]^{\nu_1+\nu_2}}\,.
\end{align}
The integral over $x$ can be evaluated in terms of the hypergeometric function
\begin{align}
\int_0^1 {dx (1-x^2)^{D/2-1} \over [2(1- x |V(y)|)]^{\nu_1+\nu_2}} = 2^{D-\nu -1}\frac{ \Gamma^2 \left(\frac{D}{2}\right)}{\Gamma (D)}
 \, _2F_1\left(\frac{\nu+1
   }{2},\frac{\nu }{2};\frac{D+1}{2} \Big| 1-y(1-y)X_{12}^2\right),
\end{align}
where $\nu=\nu_1+\nu_2$. Then, we replace the latter with its Mellin-Barnes representation
\begin{align}
\label{2MBf21}
{_{2} F_{1}}
\left.\left(
{
a, b \atop
c }
\right|
1-X
\right)
=\int \frac{dz}{2 \pi i}
X^z
\frac{\Gamma (c) \Gamma (-z) \Gamma (a + z) \Gamma (b + z) \Gamma (c - a - b - z)}{\Gamma (a) \Gamma (b) \Gamma (c-a) \Gamma (c-b)}
\, ,
\end{align} 
and combine with the previous relation to arrive at the following formula
\begin{align}\notag\label{J2}
J (\nu_1, \nu_2) {}&= {(4\pi)^{D/2}\over 4^{\nu_1+\nu_2}\Gamma(\nu_1)\Gamma(\nu_2)\Gamma(D-\nu_1-\nu_2)}
\\
{}&  \times \int{dz\over 2\pi i} (X_{12}^2/4)^{z} \Gamma(-z)\Gamma(D/2-\nu_1-\nu_2-z)\Gamma(z+\nu_1)\Gamma(z+\nu_2)
\, . 
\end{align}
Here, as before, the integration contour separates increasing and decreasing poles, generated by the product of $\Gamma-$functions, $\Gamma(-z)\Gamma(D/2-\nu_1-\nu_2-z)$ 
and $\Gamma(z+\nu_1)\Gamma(z+\nu_2)$, respectively.  According to the definition \re{J2-def}, the function $J (\nu_1,0)$ should coincide with \re{J1} for $\nu=\nu_1$. Indeed, 
for $\nu_2\to 0$ the leading contribution to \re{J2} arises from the pole at $z=0$ and it is given by \re{J1}. The relation \re{I-2pt} involves the integral \re{J2} evaluated for $\nu_1=\nu_2=(D-2)/2$.

Finally, we consider the integral containing the chain of three propagators 
\begin{align} \label{J3-def}
J (\nu_1, \nu_2, \nu_3) = \int  \frac{d\sigma_1 d\sigma_2}{|X_1 - Z_1|^{2 \nu_1} |Z_1 - Z_2|^{2 \nu_2} |Z_2 - X_2|^{2 \nu_3}} \, .
\end{align} 
Subsequently applying \re{J2}, we can express $J (\nu_1, \nu_2, \nu_3)$ as a two-fold Mellin-Barnes integral. One of the integrals can be  evaluated using the second Barnes lemma leading to
\begin{align}\notag\label{J3}
J (\nu_1, \nu_2, \nu_3)
& =\frac{(4 \pi)^{D} \Gamma (D/2 - \nu_1) \Gamma (D/2 - \nu_2) \Gamma (D/2 - \nu_3)
}{
4^{\nu_1 + \nu_2 + \nu_3} \Gamma (\nu_1) \Gamma (\nu_2)   \Gamma (\nu_3) \Gamma (D - \nu_1 - \nu_2)
\Gamma (D - \nu_1 - \nu_3) \Gamma (D - \nu_2 - \nu_3)}
\nonumber
\\[2mm]
&
\times
\int \frac{d z}{2 \pi i}
\left( X_{12}^2/{4}
\right)^z
\frac{
\Gamma (- z) \Gamma (z + \nu_1) \Gamma (z + \nu_2) \Gamma (z + \nu_3) \Gamma (D - \nu_1 - \nu_2 - \nu_3 - z)
}{
\Gamma (D/2 + z)
}
\, .
\end{align}    
As a check, we examine the limit $\nu_2\to 0$. Taking the residue at the pole $z=0$ we get
\begin{align}\notag
&
J (\nu_1, 0, \nu_3)
=
\frac{(4 \pi)^{D} \Gamma (D/2 - \nu_1)  \Gamma (D/2 - \nu_3)
}{
4^{\nu_1  + \nu_3}  \Gamma (D - \nu_1)
 \Gamma (D -   \nu_3)} =  J (\nu_1 )J (\nu_3)
 \, ,
 \end{align}     
in agreement with the expected result for \re{J3-def}. The relation \re{I-3pt} involves the integral \re{J3} evaluated for $\nu_1=\nu_2=(D-2)/2$ and $\nu_3=-j$, 
\begin{align}\label{I-J}
I(j,z) = J (1-\epsilon, 1-\epsilon, -j)\,.
\end{align}

\section{Melin-Barnes integrals}\label{app:S1}

To illustrate the technique that we use in our calculation of the the circular Wilson loop \re{W-MB}, we present detailed account for the following integral
\begin{align}\label{I-S1}
I_{S_1}=\int_0^{2\pi} ds_1 \int_0^{2\pi} ds_2 \int d\sigma_1 d\sigma_2 D_\Phi(X_1-Z_1) D_\Phi(X_2-Z_2)  |\dot X_1|  |\dot X_2| Z_{12}^2 B^2(Z_{12}^2)\,.
\end{align}
It arises in the calculation of the contribution of the diagram $S_1$ in Fig.~\ref{PolOoperatorsFig}. 

The factor of $Z_{12}^2 B^2(Z_{12}^2)$ in the right-hand side of \re{I-S1} comes from the second term in the expression for $\Pi_{S_1}$ in Eq.\ \re{tad2}. 
Following \re{Pi-MB}, we replace it with the Mellin-Barnes representation 
\begin{align}
Z_{12}^2 B^2(Z_{12}^2) = 4\int {dz_1 dz_2\over (2\pi i)^2}\, (Z_{12}^2/4)^{z_1+z_2+1} \widetilde B(z_1)  \widetilde B(z_2) \,,
\end{align}
where $\widetilde B(z_i)$ is defined in \re{A-MB1}. Substituting this relation into Eq.\ \re{I-S1}, we can perform the integrations over $Z_1$ and $Z_2$ 
using \re{I-3pt}, followed by the ones with respect to $s_1$ and $s_2$ with the help of \re{cont-int},
\begin{align}
I_{S_1}= 16\pi^{3/2} \int {dz_1 dz_2 dz_3 \over (2\pi i)^3}\,  \widetilde B(z_1)  \widetilde B(z_2)\widetilde I(z_1+z_2+1,z_3) 
\frac{\Gamma (z_3+\ft12)}{\Gamma (z_3 + 1)}\,.
\end{align}
Replacing the functions $\widetilde B$ and $\widetilde I$ by their explicit expressions, Eqs.~\re{A-MB1} and \re{tilde-I}, respectively,
we obtain the following Mellin-Barnes integral
\begin{align}\label{triple}
&
I_{S_1}
= {m^2   \sinh ^2(\pi  m) \over 2^{6-4 \epsilon} \pi ^{\frac{9}{2}-2 \epsilon }}  \int {dz_1 dz_2 dz_3\over (2\pi i)^3}
\Gamma \left(-z_1\right) \Gamma \left(-z_2\right) \Gamma \left(-z_1-2 + \epsilon \right) \Gamma \left( -z_2- 2 + \epsilon\right)
\\\notag
{}&\times \Gamma \left(2 -i m-\epsilon +z_1\right) \Gamma \left(2 + i m-\epsilon +z_1\right)
   \Gamma \left(2 -i m-\epsilon +z_2\right) \Gamma \left(2 + i m-\epsilon +z_2\right)
\\\notag
{}&\times \frac{ \Gamma \left(3 + z_1+z_2-z_3\right) \Gamma
   \left(-z_3\right) \Gamma \left(\frac{1}{2} + z_3\right) \Gamma \left(z_3-z_1-z_2-1\right)  
   \Gamma \left(3-\epsilon +z_1+z_2\right) \Gamma^2 \left(1-\epsilon +z_3\right)  }{\Gamma \left(-1 -z_1-z_2\right) \Gamma \left(1 + z_3\right) 
    \Gamma \left(2 -\epsilon +z_3\right)\Gamma^2 \left(4 -\epsilon +z_1+z_2\right)}.
\end{align}
This expression depends on the mass $m$ and the parameter of  the dimensional regularization $D=4-2\epsilon$.

The Mellin-Barnes integral \re{triple} can be analyzed using the available packages~\cite{MB}. In particular, it is straightforward to expand
$I_{S_1}$ in the Laurent series at small $\epsilon$ and compute the corresponding expansion coefficients numerically for any value of the mass $m$.
In what follows, we develop expansion of $I_{S_1}$ at small and large mass $m$.

\subsubsection*{Small mass limit}

For small $m^2$, the integral in \re{triple} is accompanied by powers of $m^2$. Therefore, $I_{S_1}$ can receive the $O(m^0)$ correction only if the  integral diverges 
as $1/m^4$ as $m\to 0$.  

Such divergences arise when the integration contour in the Mellin-Barnes integral \re{triple} is pinched by the poles. Indeed, we notice that the increasing and decreasing 
poles coming from $\Gamma \left( -z_1-2+\epsilon\right) \Gamma \left(2-i m-\epsilon +z_1\right) $ collide for $m\to 0$. As a result, the leading contribution to the integral 
in the small$-m^2$ limit only comes from the poles at $z_1=-2+\epsilon$ and $z_2=-2+\epsilon$. Evaluating the double residue, we get
\begin{align}\notag\label{S1-0}
I_{S_1}^{(0)}& = {\sqrt{\pi}
\over (4\pi)^{3-2 \epsilon}}
 {\Gamma (-1+\epsilon )\Gamma ^2(2-\epsilon )  \over   \Gamma^2 (\epsilon ) \Gamma (3-2 \epsilon )}
\\ {}&   \times
\int {dz_3\over 2\pi i}  \frac{ \Gamma \left(-z_3\right)
   \Gamma \left(z_3+\frac{1}{2}\right)  \Gamma
   \left(2 \epsilon -z_3-1\right) \Gamma \left(3-2 \epsilon +z_3\right) \Gamma^2
   \left(1-\epsilon +z_3\right)}{\Gamma \left(z_3+1\right)   \Gamma \left(-\epsilon +z_3+2\right)}\,,
   \end{align}
where we dressed it with the superscript to indicate that $I_{S_1}^{(0)}$ is the leading term in the small $m$ expansion of \re{triple}.

The factor in the first line of \re{S1-0} vanishes for $\epsilon\to 0$. Therefore, by the same reason as before, the integral is localized at poles that are pinched as $\epsilon\to 0$. 
Such poles come from $ \Gamma\left(2 \epsilon -z_3-1\right)\Gamma^2\left(-\epsilon +z_3+1\right)$. Moving the integration contour to the right, we pick up the residue at 
$z_3=-1+2\epsilon$ (with minus sign) and obtain
\begin{align}\label{LO}
I_{S_1}^{(0)}& =\frac{1}{32 \pi ^2}+O\left(\epsilon \right)\,.
\end{align}
To get the subleading correction to $I_{S_1}$ at small $m$, we rewrite $z_i-$integrals in \re{triple} as
\begin{align}
\int{d z_i\over 2\pi i} = -{\rm res}_{z_i=-2+\epsilon} + \int_{{\rm Re}\, z_i = -2+\delta}{d z_i\over 2\pi i}\,,
\end{align}
where $\delta>\epsilon$.
Here the contribution of the integral in the right-hand side is suppressed by the factor of $m^2$. Then, in the double integral over $z_1$ and $z_2$, the $O(m^2)$ correction arises from
\begin{align}\label{sep}
- \lr{\int_{{\rm Re}\,  z_1 = -2+\delta}{d z_1\over 2\pi i}\ {\rm res}_{z_2=-2+\epsilon} 
+ \int_{{\rm Re}\,  z_2 = -2+\delta}{d z_2\over 2\pi i}\  {\rm res}_{z_1=-2+\epsilon} }\,.
\end{align}
In this way, we get the $O(m^2)$ correction to \re{triple} as
\begin{align}\notag
I_{S_1}^{(1)}&=m^2\frac{\Gamma (2-\epsilon ) }{(4\pi) ^{\frac{5}{2}-2 \epsilon} }
\int{d z_1 dz_3\over (2\pi i)^2}\Gamma \left(\epsilon +z_1-z_3+1\right)
   \Gamma^2 \left(1-\epsilon +z_3\right)\Gamma \left(1-\epsilon
   -z_1+z_3\right)
\\
{}&\times
\frac{\Gamma
   \left(-z_1\right) \Gamma \left(z_1+1\right) \Gamma \left(-z_3\right) \Gamma
   \left(z_3+\frac{1}{2}\right)  \Gamma \left(\epsilon -z_1-2\right)
   \Gamma^2 \left(-\epsilon +z_1+2\right)}{\Gamma^2 \left(z_1+2\right) \Gamma \left(z_3+1\right) \Gamma
   \left(-\epsilon -z_1+1\right) \Gamma \left(-\epsilon +z_3+2\right)},
\end{align}
where the integration contour verifies ${\rm Re}\, z_1 = -2+\delta$ with $\delta>\epsilon$. 

At the next step, we examine the limit $\epsilon\to 0$. We find that, in the expressions for $I_{S_1}^{(1)}$, the increasing pole
at $z_1=-1+\epsilon$ collides with the decreasing pole at $z_1=-1$. Evaluating the residue at $z_1=-1$ we obtain
\begin{align} 
I_{S_1}^{(1)}=-\frac{m^2 }{32 \pi ^{5/2} \epsilon }\int{d z_3\over 2\pi i} \,
\Gamma^2 \left(-z_3\right) \Gamma \left(z_3+\ft{1}{2}\right) \Gamma
   \left(z_3+1\right)
+O\left(\epsilon ^0\right) = -\frac{m^2 }{16 \pi ^{2} \epsilon }+O\left(\epsilon ^0\right)\,.
\end{align}
The  $O (\epsilon^0)$ correction can be computed using the {\em MB\,Tools} package \cite{MB}. 
Being combined together with \re{LO},  this results in the small mass expansion
\begin{align} \notag
I_{S_1} = \frac{1}{32 \pi ^2} + \frac{m^2} {64 \pi ^{2}} (2\pi\e^{\gamma} )^{2\epsilon}\left[ -\frac{4}{\epsilon } -14+\frac{\pi ^2}{3}+24 \log 2 \right] + O(m^4)
\,.
\end{align}

\subsubsection*{Large mass limit}

Finally, in the large-mass limit, we replace $\sinh(\pi m)=\exp(\pi m)/2 + \dots$ and substitute the product of the $\Gamma-$functions in the second line of \re{triple} 
with its asymptotic behavior as $m \to \infty$,
\begin{align}\notag
 \Gamma(x-im)\Gamma(x+im) & = 2\pi\e^{-\pi m} m^{2x-1}\bigg[1+\frac{x(x-1)(2 x-1)}{6 m^2} 
\\[2mm]
&  +\frac{x(x-2) (x-1) (2 x-3) (2 x-1) (5 x+1)}{360 m^4} + O\left(1\over m^6\right)\bigg],
\end{align}
where $x=2-\epsilon+z$.
The resulting integral can again be evaluated using the {\em MB\,Tools} package \cite{MB} leading to
\begin{align}
I_{S_1} ={1\over 64\pi^2}\left[m^2 \left(-\frac{4}{\epsilon}+8 \log \lr{m\over\pi} -10\right)+\frac{5 \pi 
   m}{2} +\frac{5}{3} -{9\pi\over 64m} + {1\over 15m^2}+ O\left(1\over m^3\right)\right].
\end{align}
The integral develops a simple UV pole and has the following unusual feature -- it receives $O(m)$ correction. It is ultimately related to poles generated by
$\Gamma(z_3+\ft12)$ which arise from the integration over the circle, see Eq.~\re{cont-int}.

\bibliographystyle{JHEP} 

\bibliography{papers}

\providecommand{\href}[2]{#2}\begingroup\raggedright\begin{thebibliography}{10}

\bibitem{Pestun:2016zxk}
V.~Pestun et~al., \emph{{Localization techniques in quantum field theories}},
  \href{https://doi.org/10.1088/1751-8121/aa63c1}{\emph{J. Phys.} {\bfseries
  A50} (2017) 440301} [\href{https://arxiv.org/abs/1608.02952}{{\ttfamily
  1608.02952}}].

\bibitem{Pestun:2007rz}
V.~Pestun, \emph{{Localization of gauge theory on a four-sphere and
  supersymmetric Wilson loops}},
  \href{https://doi.org/10.1007/s00220-012-1485-0}{\emph{Commun. Math. Phys.}
  {\bfseries 313} (2012) 71} [\href{https://arxiv.org/abs/0712.2824}{{\ttfamily
  0712.2824}}].

\bibitem{Bobev:2013cja}
N.~Bobev, H.~Elvang, D.~Z. Freedman and S.~S. Pufu, \emph{{Holography for $N =
  2^*$ on $S^4$}}, \href{https://doi.org/10.1007/JHEP07(2014)001}{\emph{JHEP}
  {\bfseries 07} (2014) 001} [\href{https://arxiv.org/abs/1311.1508}{{\ttfamily
  1311.1508}}].

\bibitem{Andree:2010na}
R.~Andree and D.~Young, \emph{{Wilson Loops in N=2 Superconformal Yang-Mills
  Theory}}, \href{https://doi.org/10.1007/JHEP09(2010)095}{\emph{JHEP}
  {\bfseries 09} (2010) 095} [\href{https://arxiv.org/abs/1007.4923}{{\ttfamily
  1007.4923}}].

\bibitem{Billo:2019fbi}
M.~Billo, F.~Galvagno and A.~Lerda, \emph{{BPS wilson loops in generic
  conformal $ \mathcal{N} $ = 2 SU(N) SYM theories}},
  \href{https://doi.org/10.1007/JHEP08(2019)108}{\emph{JHEP} {\bfseries 08}
  (2019) 108} [\href{https://arxiv.org/abs/1906.07085}{{\ttfamily
  1906.07085}}].

\bibitem{Baggio:2014sna}
M.~Baggio, V.~Niarchos and K.~Papadodimas, \emph{{Exact correlation functions
  in $SU(2) \mathcal N=2$ superconformal QCD}},
  \href{https://doi.org/10.1103/PhysRevLett.113.251601}{\emph{Phys. Rev. Lett.}
  {\bfseries 113} (2014) 251601}
  [\href{https://arxiv.org/abs/1409.4217}{{\ttfamily 1409.4217}}].

\bibitem{Baggio:2014ioa}
M.~Baggio, V.~Niarchos and K.~Papadodimas, \emph{{tt$^{*}$ equations,
  localization and exact chiral rings in 4d $ \mathcal{N} $ =2 SCFTs}},
  \href{https://doi.org/10.1007/JHEP02(2015)122}{\emph{JHEP} {\bfseries 02}
  (2015) 122} [\href{https://arxiv.org/abs/1409.4212}{{\ttfamily 1409.4212}}].

\bibitem{Baggio:2015vxa}
M.~Baggio, V.~Niarchos and K.~Papadodimas, \emph{{On exact correlation
  functions in SU(N) $ \mathcal{N}=2 $ superconformal QCD}},
  \href{https://doi.org/10.1007/JHEP11(2015)198}{\emph{JHEP} {\bfseries 11}
  (2015) 198} [\href{https://arxiv.org/abs/1508.03077}{{\ttfamily
  1508.03077}}].

\bibitem{Gerchkovitz:2016gxx}
E.~Gerchkovitz, J.~Gomis, N.~Ishtiaque, A.~Karasik, Z.~Komargodski and S.~S.
  Pufu, \emph{{Correlation Functions of Coulomb Branch Operators}},
  \href{https://doi.org/10.1007/JHEP01(2017)103}{\emph{JHEP} {\bfseries 01}
  (2017) 103} [\href{https://arxiv.org/abs/1602.05971}{{\ttfamily
  1602.05971}}].

\bibitem{Baggio:2016skg}
M.~Baggio, V.~Niarchos, K.~Papadodimas and G.~Vos, \emph{{Large-N correlation
  functions in $ \mathcal{N} $ = 2 superconformal QCD}},
  \href{https://doi.org/10.1007/JHEP01(2017)101}{\emph{JHEP} {\bfseries 01}
  (2017) 101} [\href{https://arxiv.org/abs/1610.07612}{{\ttfamily
  1610.07612}}].

\bibitem{Rodriguez-Gomez:2016ijh}
D.~Rodriguez-Gomez and J.~G. Russo, \emph{{Large N Correlation Functions in
  Superconformal Field Theories}},
  \href{https://doi.org/10.1007/JHEP06(2016)109}{\emph{JHEP} {\bfseries 06}
  (2016) 109} [\href{https://arxiv.org/abs/1604.07416}{{\ttfamily
  1604.07416}}].

\bibitem{Rodriguez-Gomez:2016cem}
D.~Rodriguez-Gomez and J.~G. Russo, \emph{{Operator mixing in large $N$
  superconformal field theories on S$^{4}$ and correlators with Wilson loops}},
  \href{https://doi.org/10.1007/JHEP12(2016)120}{\emph{JHEP} {\bfseries 12}
  (2016) 120} [\href{https://arxiv.org/abs/1607.07878}{{\ttfamily
  1607.07878}}].

\bibitem{Billo:2017glv}
M.~Billo, F.~Fucito, A.~Lerda, J.~F. Morales, {\relax Ya}.~S. Stanev and
  C.~Wen, \emph{{Two-point Correlators in N=2 Gauge Theories}},
  \href{https://doi.org/10.1016/j.nuclphysb.2017.11.003}{\emph{Nucl. Phys.}
  {\bfseries B926} (2018) 427}
  [\href{https://arxiv.org/abs/1705.02909}{{\ttfamily 1705.02909}}].

\bibitem{Billo:2018oog}
M.~Billo, F.~Galvagno, P.~Gregori and A.~Lerda, \emph{{Correlators between
  Wilson loop and chiral operators in $ \mathcal{N}=2 $ conformal gauge
  theories}}, \href{https://doi.org/10.1007/JHEP03(2018)193}{\emph{JHEP}
  {\bfseries 03} (2018) 193}
  [\href{https://arxiv.org/abs/1802.09813}{{\ttfamily 1802.09813}}].

\bibitem{Billo:2019job}
M.~Billo, F.~Fucito, G.~P. Korchemsky, A.~Lerda and J.~F. Morales,
  \emph{{Two-point correlators in non-conformal $ \mathcal{N} $ = 2 gauge
  theories}}, \href{https://doi.org/10.1007/JHEP05(2019)199}{\emph{JHEP}
  {\bfseries 05} (2019) 199}
  [\href{https://arxiv.org/abs/1901.09693}{{\ttfamily 1901.09693}}].

\bibitem{Adler:1972qq}
S.~L. Adler, \emph{{Massless, Euclidean Quantum Electrodynamics on the
  Five-Dimensional Unit Hypersphere}},
  \href{https://doi.org/10.1103/physrevd.7.3821.2,
  10.1103/PhysRevD.6.3445}{\emph{Phys. Rev.} {\bfseries D6} (1972) 3445}.

\bibitem{Adler:1973ty}
S.~L. Adler, \emph{{Massless Electrodynamics on the Five-Dimensional Unit
  Hypersphere: an Amplitude - Integral Formulation}},
  \href{https://doi.org/10.1103/PhysRevD.8.2400,
  10.1103/PhysRevD.15.1803}{\emph{Phys. Rev.} {\bfseries D8} (1973) 2400}.

\bibitem{Drummond:1975yc}
I.~T. Drummond, \emph{{Dimensional Regularization of Massless Theories in
  Spherical Space-Time}},
  \href{https://doi.org/10.1016/0550-3213(75)90089-9}{\emph{Nucl. Phys.}
  {\bfseries B94} (1975) 115}.

\bibitem{Drummond:1977uy}
I.~T. Drummond and G.~M. Shore, \emph{{Dimensional Regularization of Massless
  Quantum Electrodynamics in Spherical Space-Time. 1.}},
  \href{https://doi.org/10.1016/0003-4916(79)90045-9}{\emph{Annals Phys.}
  {\bfseries 117} (1979) 89}.

\bibitem{Belitsky:2000ii}
A.~V. Belitsky, S.~Vandoren and P.~van Nieuwenhuizen, \emph{{Instantons,
  Euclidean supersymmetry and Wick rotations}},
  \href{https://doi.org/10.1016/S0370-2693(00)00183-0}{\emph{Phys. Lett.}
  {\bfseries B477} (2000) 335}
  [\href{https://arxiv.org/abs/hep-th/0001010}{{\ttfamily hep-th/0001010}}].

\bibitem{Buchel:2000cn}
A.~Buchel, A.~W. Peet and J.~Polchinski, \emph{{Gauge dual and noncommutative
  extension of an N=2 supergravity solution}},
  \href{https://doi.org/10.1103/PhysRevD.63.044009}{\emph{Phys. Rev.}
  {\bfseries D63} (2001) 044009}
  [\href{https://arxiv.org/abs/hep-th/0008076}{{\ttfamily hep-th/0008076}}].

\bibitem{Festuccia:2011ws}
G.~Festuccia and N.~Seiberg, \emph{{Rigid Supersymmetric Theories in Curved
  Superspace}}, \href{https://doi.org/10.1007/JHEP06(2011)114}{\emph{JHEP}
  {\bfseries 06} (2011) 114} [\href{https://arxiv.org/abs/1105.0689}{{\ttfamily
  1105.0689}}].

\bibitem{Burges:1985qq}
C.~J.~C. Burges, D.~Z. Freedman, S.~Davis and G.~W. Gibbons,
  \emph{{Supersymmetry in Anti-de Sitter Space}},
  \href{https://doi.org/10.1016/0003-4916(86)90203-4}{\emph{Annals Phys.}
  {\bfseries 167} (1986) 285}.

\bibitem{Maldacena:1998im}
J.~M. Maldacena, \emph{{Wilson loops in large N field theories}},
  \href{https://doi.org/10.1103/PhysRevLett.80.4859}{\emph{Phys. Rev. Lett.}
  {\bfseries 80} (1998) 4859}
  [\href{https://arxiv.org/abs/hep-th/9803002}{{\ttfamily hep-th/9803002}}].

\bibitem{Erickson:2000af}
J.~K. Erickson, G.~W. Semenoff and K.~Zarembo, \emph{{Wilson loops in N=4
  supersymmetric Yang-Mills theory}},
  \href{https://doi.org/10.1016/S0550-3213(00)00300-X}{\emph{Nucl. Phys.}
  {\bfseries B582} (2000) 155}
  [\href{https://arxiv.org/abs/hep-th/0003055}{{\ttfamily hep-th/0003055}}].

\bibitem{Drukker:2000rr}
N.~Drukker and D.~J. Gross, \emph{{An Exact prediction of N=4 SUSYM theory for
  string theory}}, \href{https://doi.org/10.1063/1.1372177}{\emph{J. Math.
  Phys.} {\bfseries 42} (2001) 2896}
  [\href{https://arxiv.org/abs/hep-th/0010274}{{\ttfamily hep-th/0010274}}].

\bibitem{Plefka:2001bu}
J.~Plefka and M.~Staudacher, \emph{{Two loops to two loops in N=4
  supersymmetric Yang-Mills theory}},
  \href{https://doi.org/10.1088/1126-6708/2001/09/031}{\emph{JHEP} {\bfseries
  09} (2001) 031} [\href{https://arxiv.org/abs/hep-th/0108182}{{\ttfamily
  hep-th/0108182}}].

\bibitem{Arutyunov:2001hs}
G.~Arutyunov, J.~Plefka and M.~Staudacher, \emph{{Limiting geometries of two
  circular Maldacena-Wilson loop operators}},
  \href{https://doi.org/10.1088/1126-6708/2001/12/014}{\emph{JHEP} {\bfseries
  12} (2001) 014} [\href{https://arxiv.org/abs/hep-th/0111290}{{\ttfamily
  hep-th/0111290}}].

\bibitem{Gatheral:1983cz}
J.~G.~M. Gatheral, \emph{{Exponentiation of Eikonal Cross-sections in
  Nonabelian Gauge Theories}},
  \href{https://doi.org/10.1016/0370-2693(83)90112-0}{\emph{Phys. Lett.}
  {\bfseries 133B} (1983) 90}.

\bibitem{Frenkel:1984pz}
J.~Frenkel and J.~C. Taylor, \emph{{Nonabelian eikonal exponentiation}},
  \href{https://doi.org/10.1016/0550-3213(84)90294-3}{\emph{Nucl. Phys.}
  {\bfseries B246} (1984) 231}.

\bibitem{unpublished1}
M.~Billo, F.~Fucito, A.~Lerda and J.~F. Morales, \emph{{unpublished}},  2019.

\bibitem{unpublished2}
M.~Bonini, L.~Griguolo, M.~Preti and D.~Seminara, \emph{{unpublished}},  2019.

\bibitem{MB}
\emph{MB Tools}. {\tt https://mbtools.hepforge.org/}.

\bibitem{Smirnov:2012gma}
V.~A. Smirnov, \emph{{Analytic tools for Feynman integrals}},
  \href{https://doi.org/10.1007/978-3-642-34886-0}{\emph{Springer Tracts Mod.
  Phys.} {\bfseries 250} (2012) 1}.

\bibitem{Chernikov:1968zm}
N.~A. Chernikov and E.~A. Tagirov, \emph{{Quantum theory of scalar fields in de
  Sitter space-time}}, {\emph{Ann. Inst. H. Poincare Phys. Theor.} {\bfseries
  A9} (1968) 109}.

\bibitem{Dowker:1975tf}
J.~S. Dowker and R.~Critchley, \emph{{Effective Lagrangian and Energy Momentum
  Tensor in de Sitter Space}},
  \href{https://doi.org/10.1103/PhysRevD.13.3224}{\emph{Phys. Rev.} {\bfseries
  D13} (1976) 3224}.

\bibitem{Candelas:1975du}
P.~Candelas and D.~J. Raine, \emph{{General Relativistic Quantum Field
  Theory-An Exactly Soluble Model}},
  \href{https://doi.org/10.1103/PhysRevD.12.965}{\emph{Phys. Rev.} {\bfseries
  D12} (1975) 965}.

\end{thebibliography}\endgroup

\end{document}